\documentclass[acmsmall,screen]{acmart}
\usepackage{amsmath}
\usepackage{textcomp}
\usepackage{algorithm}
\usepackage{algpseudocode}
\usepackage{float}  
\usepackage{stfloats}
\usepackage{lipsum}  
\usepackage{multirow}
\usepackage{caption} 
\usepackage{tabularx} 
\usepackage{placeins}
\usepackage{makecell} 
\usepackage{graphicx} 
\usepackage{array}     
\usepackage{xcolor}
\usepackage{pifont}
\usepackage{booktabs}
\usepackage{adjustbox}
\usepackage{hyperref} 

\newcommand{\cmark}{\ding{51}} % Check mark
\newcommand{\xmark}{\ding{55}} % Cross mark

\newcolumntype{C}[1]{>{\centering\arraybackslash}m{#1}} 
\newcolumntype{L}[1]{>{\centering\arraybackslash}m{#1}} 

\AtBeginDocument{%
  }

\setcopyright{acmlicensed}    
\copyrightyear{2026}
\acmVolume{X}              
\acmNumber{X}                 
\acmArticle{XXX}             
\acmMonth{8}                 
\acmDOI{10.1145/XXXXXXX.XXXXXXX} 


\begin{document}
\title{VARA: A Voltage-Aware ReRAM-Based Accelerator for Energy-Efficient Computing}

\author{Peng Dang}
\orcid{0009-0008-9409-2798}
\affiliation{%
  \institution{SKLP, Institute of Computing Technology, Chinese Academy of Sciences}
  \city{Beijing}
  \country{China}
}
\affiliation{%
  \institution{University of Chinese Academy of Sciences}
  \city{Beijing}
  \country{China}
}
\email{dangpeng21@mails.ucas.ac.cn}

\author{Yintao He}
\orcid{0009-0008-9409-2798}
\affiliation{%
  \institution{SKLP, Institute of Computing Technology, Chinese Academy of Sciences}
  \city{Beijing}
  \country{China}
}
\affiliation{%
  \institution{University of Chinese Academy of Sciences}
  \city{Beijing}
  \country{China}
}

\author{Huawei Li}
\orcid{0009-0008-9409-2798}
% \authornote{Corresponding author.}
\affiliation{%
  \institution{SKLP, Institute of Computing Technology, Chinese Academy of Sciences}
  \city{Beijing}
  \country{China}
}
\affiliation{%
  \institution{University of Chinese Academy of Sciences}
  \city{Beijing}
  \country{China}
}
\email{lihuawei@ict.ac.cn}
\renewcommand{\shortauthors}{Peng Dang et al.}

\begin{abstract}
\makebox[\textwidth][c]{\rule{\textwidth}{0.4pt}}
ReRAM-based in-memory computing (IMC) architectures are widely regarded as a promising approach to alleviating the computational bottleneck of conventional architectures. Since ReRAM crossbars perform matrix-vector multiplication (MVM) in the analog domain, their computational energy consumption is highly dependent on weight and activation distributions. However, most existing ReRAM accelerators focus primarily on weight optimization while paying limited attention to the impact of activations on computational energy consumption, leaving the energy-saving potential of activation sparsity largely underexploited. In this paper, we propose a voltage-aware ReRAM-based accelerator (VARA), along with its accompanying design methodology. Specifically, we first introduce a voltage-aware training (VAT) algorithm that incorporates a preset threshold into the activation function to steer the activation distribution toward zero values, thereby enhancing activation sparsity. Building upon this, we further propose a co-zero activation reordering (CAR) scheme for crossbar-level computation skipping. CAR clusters activation dimensions based on their co-zero correlations and consistently reorders both the activation matrix and its corresponding weights. This process consolidates scattered zero activations into contiguous zero-valued regions to maximize the benefits of crossbar-level computation skipping. Extensive experimental results demonstrate that, with only marginal accuracy loss, VARA reduces the average total system energy consumption by 60.12\% and improves the average system energy efficiency by 2.68$\times$ compared to the baseline, outperforming existing state-of-the-art accelerators for sparse-activation optimization.
\end{abstract}


\maketitle
\section{Introduction}
\label{sec:section_1}

In recent years, deep neural networks (DNNs) have been widely applied in domains such as computer vision and natural language processing~\cite{mehonic2022brain}. However, as model sizes and computational complexity continue to increase, deploying DNNs on resource-constrained edge devices poses significant challenges in terms of energy efficiency~\cite{zhang2024algorithm,3589062}. To address these challenges, researchers have extensively explored hardware acceleration, model compression, and emerging computing paradigms. These efforts aim to improve the computational efficiency of neural networks by reducing computational and storage overheads~\cite{song2021brahms}. Among these approaches, resistive random-access memory (ReRAM)-based IMC architectures integrate storage and computation, enabling MVM to be performed directly within crossbar arrays. This integration substantially reduces data movement between memory and processing units~\cite{zhang2019design}. As a result, ReRAM-based neural network accelerators are considered a promising approach to alleviating the ``memory wall'' bottleneck inherent in conventional von Neumann architectures~\cite{9716051,liu202033,sebastian2020memory}.

In ReRAM-based accelerators, MVM operations are performed in parallel within crossbars in the analog domain. Their computational energy consumption depends strongly on device conductances and input voltages. Specifically, neural network weights are mapped to the conductances of ReRAM cells, while input activations are encoded as voltages applied to the crossbar. The input voltages and device conductances jointly determine the currents generated by individual memory cells. These currents are subsequently accumulated along the bitlines to produce analog output currents~\cite{zhang2019design}. The magnitudes of the bitline currents affect not only the energy consumption of the crossbar itself but also that of the downstream peripheral circuits~\cite{he2020towards}. Consequently, the value distributions of both neural network weights and activations significantly influence the overall computational energy consumption of ReRAM accelerators by altering the internal crossbar currents.

Prior studies~\cite{he2020towards,he2021saving,3576195} have demonstrated that the weight distributions of binary neural networks (BNNs) can be shaped to increase the proportion of high-resistance-state (HRS) devices in ReRAM crossbars. A higher HRS proportion suppresses bitline currents and reduces the power consumption of crossbars and analog-to-digital converters (ADCs).
However, these approaches primarily focus on optimizing weight distributions and encoding schemes, while paying limited attention to the impact of activation distributions on system energy consumption. In fact, activations are mapped to input voltages in ReRAM accelerators, and their value distributions likewise directly affect the magnitudes of bitline currents within the crossbars.

\begin{figure}[!t]
    \centering
    \vspace{3pt}
    \includegraphics[width=0.6\linewidth]{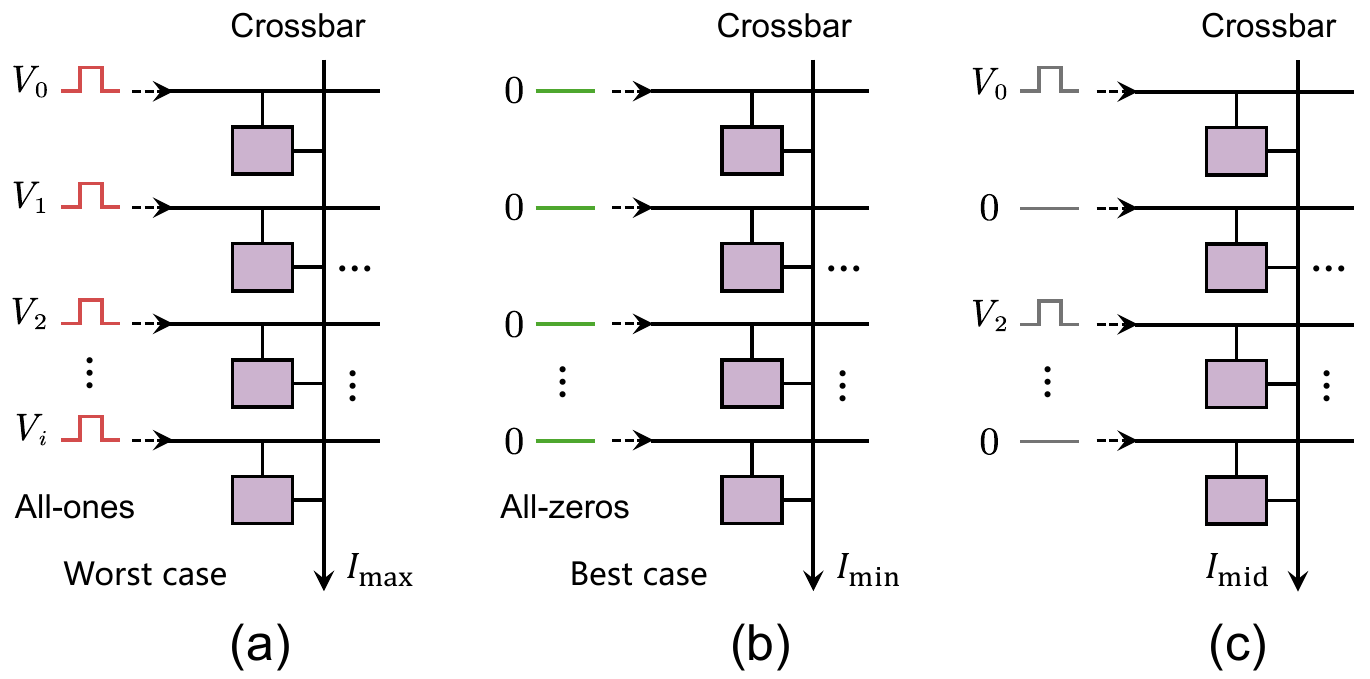}
    \vspace{-8pt}
    \caption{Impact of activation distributions on the output current of ReRAM crossbars.}
    \Description{None}
    \vspace{-6pt}
    \label{fig:Figure1}
\end{figure}

As illustrated in Fig.~\ref{fig:Figure1}(a), when all input activations are ``1'', valid read voltages are applied to all wordlines. The bitline current consequently reaches its maximum value ($I_{\text{max}}$). Under this condition, the ReRAM crossbar operates at peak power consumption ($P_{\text{max}}$), and the downstream ADC also operates at its maximum power level. Conversely, as shown in Fig.~\ref{fig:Figure1}(b), when all input activations are ``0'', the corresponding wordline voltages become zero. No effective multiply-accumulate (MAC) operations occur within the crossbar, and the bitline current drops to zero. This phenomenon indicates that the activation distribution significantly affects the power consumption of both the crossbar and its peripheral circuits, particularly the energy-intensive ADC modules. Furthermore, when a crossbar receives an all-zero input vector, its output is necessarily zero. Therefore, the associated crossbar computation and analog-to-digital conversion are unnecessary. This computational predictability enables the system to selectively skip these redundant operations and eliminate unnecessary dynamic power consumption. This activation-dependent computational characteristic provides a key design insight for the energy-efficiency optimization strategy proposed in this work.

Although the all-zero input skipping mechanism holds substantial potential for energy efficiency optimization, existing studies~\cite{9586315,10323699,yang2019sparse,yuan2021forms,10177200,3287640} still face two fundamental limitations. First, although conventional activation functions such as ReLU naturally produce a certain proportion of zero activations, standard model training does not explicitly optimize activation sparsity, imposing an upper bound on the proportion of exploitable all-zero inputs. Second, and more importantly, high activation sparsity does not directly translate into opportunities for crossbar computation skipping. Due to the highly scattered distribution of nonzero activations across input dimensions, even when the overall activation matrix is extremely sparse, the input vectors mapped to individual ReRAM crossbars are likely to contain a few nonzero elements. This forces the corresponding crossbars and peripheral circuits to execute full computations, preventing the triggering of all-zero input skipping. Therefore, simultaneously increasing the proportion of zero activations and transforming scattered zero activations into directly exploitable all-zero input vectors remains a key challenge in maximizing the benefits of crossbar-level computation skipping.

To address the aforementioned challenges, this paper proposes VARA, a voltage-aware ReRAM accelerator architecture. By integrating the VAT algorithm with the CAR scheme, VARA enables co-optimization across model training and deployment. Specifically, the main contributions of this work are as follows:

\begin{itemize}
\item We propose VARA, a voltage-aware ReRAM accelerator architecture that exploits the inherent relationship between activation distributions and computational power consumption. Through co-design spanning model training and deployment, VARA effectively harnesses the energy-efficiency potential of activation sparsity.

\item We propose the VAT algorithm to enhance activation sparsity. By introducing a predefined threshold into the activation function and jointly optimizing network parameters during training, VAT drives low-magnitude activations toward zero, significantly increasing the zero activation ratio (ZAR) without compromising model accuracy.

\item Building upon the sparsity model induced by the VAT algorithm, we propose a CAR scheme. While preserving computational equivalence, CAR reorders the dimensions of activation and weight matrices to consolidate scattered zero activations into contiguous structured sparse regions, maximizing the occurrence probability of all-zero input vectors.

\item Experimental results across multiple networks and datasets demonstrate that, compared with the baseline, VARA reduces total system energy consumption by an average of 60.12\% and improves system energy efficiency by an average of 2.68$\times$, while incurring only a minor accuracy loss. These results validate the effectiveness of the proposed design in improving the energy efficiency of ReRAM accelerators.

\end{itemize}

The remainder of this paper is organized as follows. Section~\ref{sec:section_2} introduces the background of ReRAM-based accelerators and analyzes the relationship between activation distributions and system-level computational energy consumption. Section~\ref{sec:section_3} describes the proposed VARA architecture, detailing the co-optimization mechanism of VAT and CAR. Section~\ref{sec:section_4} presents the experimental setup and evaluation results, along with comparisons against baseline designs. Finally, Section~\ref{sec:section_5} concludes the paper.
\section{Background and Motivation}
\label{sec:section_2}
\subsection{ReRAM-Based Neural Network Accelerators}
\label{subsec}

\begin{figure*}[!t]
    \centering
    \includegraphics[width=0.8\linewidth]{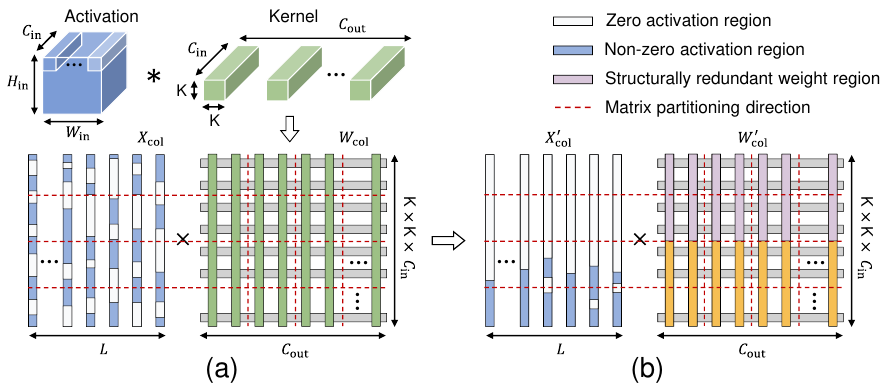}
    \vspace{-10pt}
    \caption{Comparison between conventional matrix computation and co-zero activation reordering. (a) In the conventional implementation, zero activations are scattered, making crossbar-level skipping difficult. (b) Co-zero activation reordering consistently reorders the activation and weight, consolidating scattered zero activations into highly concentrated zero regions and creating skippable redundant weight regions.}
    \Description{None}
    \vspace{-8pt}
    \label{fig:Figure2} 
\end{figure*}

In neural network computing, MVM constitutes the dominant computational workload~\cite{9731725,10375354}. By integrating storage and computation, ReRAM-based accelerators can efficiently execute MVM operations within crossbar arrays~\cite{yan2023improving,geng2021chip,song2017pipelayer}. However, since ReRAM crossbars natively support only MVM operations, extending their in-situ parallel computing capability to convolutional layers requires the image-to-column (Img2Col) transformation. This process converts convolutions into general matrix multiplication (GEMM), which is subsequently decomposed into a series of MVM operations executable by the crossbar~\cite{7551379,3304049}. As illustrated in Fig.~\ref{fig:Figure2}(a), let the spatial dimensions of the input activation feature map be $H_{\text{in}}\times W_{\text{in}}$, the number of input channels be $C_{\text{in}}$, the kernel size be $K\times K$, and the number of output channels be $C_{\text{out}}$. Through the Img2Col operation, the input feature map is unrolled via a sliding window into a two-dimensional activation matrix $X_{\text{col}}$, while the convolution kernels are reshaped into a weight matrix $W_{\text{col}}$. Following this transformation, the original convolution is equivalently formulated as a standard matrix multiplication $Y_{\text{col}} = W_{\text{col}} X_{\text{col}}$. This equivalent formulation allows convolutional computation to be mapped onto ReRAM crossbars in matrix form~\cite{3358328}.

As illustrated in Fig.~\ref{fig:Figure1}, in ReRAM accelerators, neural network weights are mapped to the conductance values of ReRAM cells, while input activations are converted into analog voltages via digital-to-analog converters (DACs) and applied to the crossbar along the wordlines~\cite{zhang2019design}. According to Ohm’s law and Kirchhoff’s current law, each input voltage is multiplied by the corresponding cell conductance, and the resulting currents are accumulated on the bitlines to form the output current~\cite{11100389}. Specifically, the output current on the $j$-th bitline can be expressed as:
\begin{equation}
I_j = \sum_{i=1}^{N} G_{ij} \cdot V_i
\label{eq:I_j}
\end{equation}
where $G_{ij}$ denotes the conductance of the ReRAM cell at the $i$-th row and $j$-th column, $V_i$ represents the input voltage applied to the $i$-th wordline, and $I_j$ indicates the output current on the $j$-th bitline. Subsequently, the bitline currents are quantized into digital signals by ADCs and forwarded to downstream processing units for further computation~\cite{10753271}.

During model deployment, a single ReRAM crossbar cannot directly accommodate an entire MVM task due to its limited physical dimensions, interconnect IR drop, and the drive capability of peripheral circuits. Consequently, the complete MVM computation must be partitioned into multiple subtasks and mapped across several ReRAM crossbars for parallel processing~\cite{sun2018xnor,9218657}. The analog partial sum currents from each crossbar are quantized by ADCs and then globally accumulated in the digital domain to reconstruct the final MVM result. However, this mixed-signal computing architecture introduces new design challenges. Specifically, high-resolution ADCs required to ensure readout accuracy of analog partial sums incur substantial power overhead, becoming a critical bottleneck that limits further improvements in system energy efficiency~\cite{9964075,10044806,9138916}.

\subsection{Activation-Dependent Power Characteristics in ReRAM Accelerators}
\label{subsec:activation}

At the crossbar input, input activations are converted into analog voltages $V_i$ via DACs and applied to the corresponding wordlines of the crossbar. During analog computation, the power consumption associated with the $j$-th bitline can be approximated as
\begin{equation}
P_{\text{XB},j} = \sum_i V_i^2 G_{ij}
= \sum_i I_{ij}^{2} R_{ij}
\end{equation}
This expression indicates that when an activation value is zero, the voltage applied to the corresponding wordline becomes zero, and the dynamic power consumption of the associated memory cell also drops to zero. Therefore, increasing the proportion of zero activations can directly reduce the dynamic energy consumption of ReRAM crossbars.

At the crossbar output, bitline currents also affect the dynamic energy consumption of the subsequent ADCs~\cite{he2020towards,he2021saving,3576195}. As a major contributor to peripheral power consumption in ReRAM accelerators, an ADC consumes dynamic power primarily through the charging and discharging of its sampling capacitor~\cite{7244259}. Letting $R_L$ denote the load resistance, the relationship between the ADC input voltage $V_{\text{in}}$ and the current $I_j$ on the $j$-th bitline is given by $V_{\text{in}} = I_j R_L$. At a sampling frequency of $f$, the dynamic power consumption of the ADC in the $j$-th column can be approximated as~\cite{zhang2019design,7244259}
\begin{equation}
P_{\text{ADC},j} = \frac{1}{2} f C V_{\text{in}}^{2}
= \frac{1}{2} f C \bigl(I_j R_L\bigr)^{2}
\label{eq:W_sample}
\end{equation}
where $C$ represents the sampling capacitance. Combining Eq.~(\ref{eq:I_j}) and Eq.~(\ref{eq:W_sample}) reveals that, for fixed hardware parameters $f$, $C$, and $R_L$, the ADC dynamic power is proportional to the square of the bitline current, which is jointly determined by input voltages and weight conductances. Consequently, optimizing the activation distribution not only reduces the computational power consumption of ReRAM crossbars but also exploits the quadratic dependence of ADC dynamic power on bitline current to achieve substantial energy savings.

\subsection{Research Motivation}
\label{subsec:motivation}

As discussed in Section~\ref{sec:section_1}, when the input vector to a ReRAM crossbar is entirely zero, the crossbar output can be predicted to be zero before computation. By exploiting this computational predictability, the system can directly skip the corresponding crossbar-level computation. Such hardware-level optimization potential can be effectively unlocked through a hardware-aware training strategy. The significant over-parameterization of DNNs provides considerable flexibility for regulating their internal activation distributions~\cite{zhang2016understanding}. Prior studies~\cite{he2020towards,he2021saving,3576195} have demonstrated that introducing a predefined threshold into the sign function can effectively reshape the weight distributions of BNNs. Inspired by this observation, we modify the zero-decision threshold of nonlinear activation functions to actively drive low-magnitude activations that contribute little to network representations toward zero. Regulating activation distributions in this manner not only substantially increases the ZAR but also establishes the sparsity foundation for crossbar-level computation skipping.

However, high element-wise activation sparsity alone does not necessarily translate into increased opportunities for crossbar-level computation skipping. As shown in Fig.~\ref{fig:Figure2}(a), conventional matrix mapping typically partitions along the activation dimension order without considering co-zero correlations among different activation dimensions. Even when the activation matrix is globally sparse, zero activations scattered across input dimensions result in nonzero elements within individual crossbar input vectors, thereby preventing crossbar-level skipping. Furthermore, sorting by the independent zero rate of each dimension is insufficient, as it captures only the marginal zero probability per dimension while ignoring the joint distribution of dimensions that are simultaneously zero at identical input positions. Therefore, it is essential to exploit co-zero correlations among activation dimensions during the mapping phase to group dimensions with similar zero patterns into the same crossbar, as illustrated in Fig.~\ref{fig:Figure2}(b). Through such reordering, the probability of all-zero input vectors is maximized, converting discrete element-wise sparsity into substantial crossbar redundancy, thereby maximizing the benefits of computation skipping.

Based on the above analysis, this work identifies the activation distribution as the primary target for energy-efficiency optimization in ReRAM accelerators. The feasibility of this approach rests on two interrelated observations. First, the activation distribution can be actively regulated during training to increase the proportion of zero activations. Second, co-zero correlations among activation dimensions can be leveraged during deployment to reorder and consolidate scattered zero activations. This process transforms element-wise sparsity into exploitable structured redundancy. To this end, we propose a VAT algorithm and a CAR scheme. Specifically, VAT increases the element-wise ZAR by adjusting the zero-decision threshold of the activation function. CAR then consistently reorders the activation dimensions and their corresponding weight mappings, consolidating scattered zero activations into contiguous, structured all-zero regions. This reordering substantially increases the probability of triggering hardware-level computation skipping. Together, VAT and CAR enable joint optimization of system-level performance.

\section{Methodology}
\label{sec:section_3}
This section systematically presents the proposed VARA architecture. We first provide an overview of the overall architectural design of VARA. Subsequently, we detail the VAT algorithm for reshaping activation distributions. Finally, we elaborate on the CAR scheme and explain how it transforms the sparsity induced by VAT into exploitable structured redundancy to reduce system energy consumption.

\subsection{VARA Architecture}
We propose VARA, a voltage-aware ReRAM accelerator architecture. As illustrated in Fig.~\ref{fig:Figure2}, VARA adopts a hierarchical organization and primarily consists of buffers, index units (IUs), data conversion modules, accumulation units, and processing elements (PEs). The buffers store input data and intermediate computation results. The IUs store the activation-dimension permutation indices generated by the co-zero reordering algorithm. The DACs convert activations into analog input voltages, while the ADCs convert bitline output currents into digital signals. The accumulation units aggregate the partial sums produced by individual crossbars to generate the final computation results. Each PE performs the core MVM operations. Multiple PEs are further integrated into a tile, and the tiles are interconnected through a network-on-chip (NoC), forming a scalable parallel computing architecture.

\begin{figure}[!t]
    \centering
    \includegraphics[width=1\linewidth]{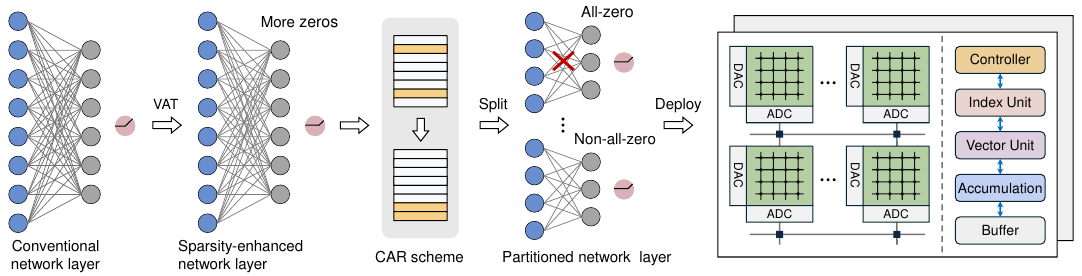}
    \vspace{-15pt}
    \caption{The VARA architecture based on the VAT algorithm and CAR scheme.}
    \Description{None}
    \vspace{-12pt}
    \label{fig:Figure3}
\end{figure}

As shown in Fig.~\ref{fig:Figure3}, VARA integrates activation distribution optimization, crossbar mapping, and runtime computation skipping through the coordinated use of VAT and CAR. During training, VAT adjusts the zero-decision thresholds of activation functions to increase the proportion of zero activations, thereby reducing the crossbar computation overhead associated with nonzero inputs. However, nonzero activations dispersed across the dimensions of an activation matrix make all-zero input blocks less likely to occur, limiting computation-skipping opportunities. To exploit zero activations more effectively, we propose CAR, which reorders input dimensions according to co-zero relationships among activation dimensions. This increases the proportion of all-zero input blocks aligned with the crossbar input capacity, creating opportunities for crossbar-level computation skipping.

CAR combines offline profiling with online inference. During offline profiling, co-zero correlations between input dimensions are computed from the activations produced by the VAT-pretrained model over the entire training set. These correlations are used to generate and freeze permutation indices, which are stored in the IU. The corresponding dimensions of the weight matrix are then reordered using these indices, and the reordered matrix is partitioned according to the crossbar dimensions. The resulting weight submatrices are programmed into ReRAM crossbars. During online inference, the system reorders activations at runtime using the fixed indices and partitions the reordered activation matrix into input blocks. Zero-block detection then determines which computations can be skipped: crossbar computations and associated peripheral readout operations are bypassed for all-zero blocks, whereas blocks containing nonzero activations are processed normally. By optimizing the ordering of input dimensions offline, CAR increases the likelihood of forming all-zero input blocks during inference, thereby creating more opportunities for runtime computation skipping.

\subsection{Voltage-Aware Training Algorithm}
\subsubsection{Voltage-Aware Training Algorithm for Step Functions}
\begin{figure}[t]
    \centering
    \includegraphics[width=1.0\linewidth]{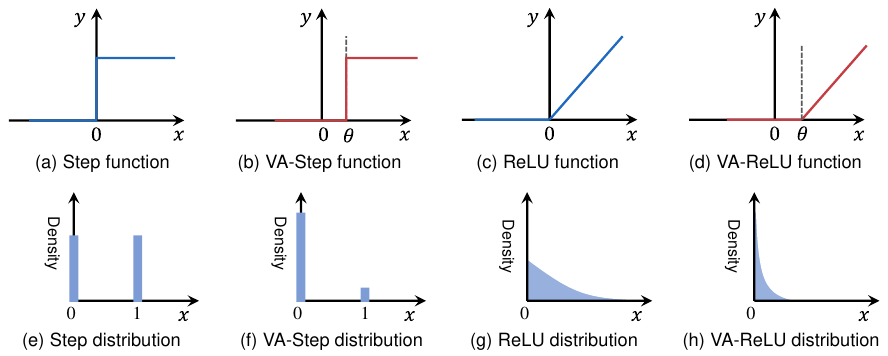}
    \vspace{-15pt}
    \caption{Comparison of activation functions. (a)--(d) show the Step, VA-Step, ReLU, and VA-ReLU functions, respectively; (e)--(h) depict their corresponding activation distributions.}
    \Description{None}
    \vspace{-8pt}
    \label{fig:Figure4}
\end{figure}

In conventional BNN implementations~\cite{3576195,9157443,xu2021recu}, the sign function (Sign) is typically used during forward propagation to constrain real-valued activations to $\{-1, +1\}$. In this work, to enable zero activations to be mapped to zero input voltages, we adopt a step function (Step), as illustrated in Fig.~\ref{fig:Figure4}(a), to restrict activations to $\{0, +1\}$. Furthermore, to enhance activation sparsity, we introduce an adjustable activation threshold $\theta$ into the activation binarization process, constructing a voltage-aware step (VA-Step) function, as shown in Fig.~\ref{fig:Figure4}(b). Given a real-valued pre-activation $a_r$, the binarization process is defined as follows:
\begin{equation}
a_b = 
\begin{cases} 
     +1, & \text{if } a_r \geq \theta, \\
     0,  & \text{otherwise}.
\end{cases}
\label{eq:a_b}
\end{equation}
where $a_b$ denotes the binarized activation output.

\begin{figure}[t]
    \centering
    \includegraphics[width=0.7\linewidth]{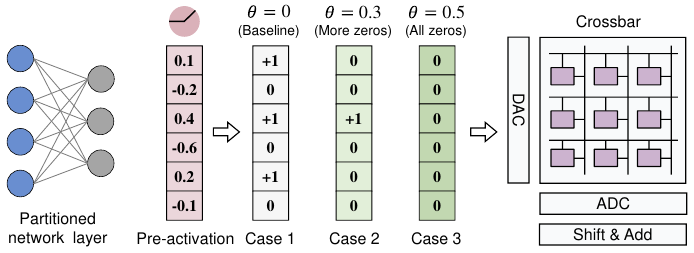}
    \vspace{-3pt}
    \caption{Activation outputs of VA-Step under different activation thresholds.}
    \Description{None}
    \vspace{-8pt}
    \label{fig:Figure5}
\end{figure}

As shown in Fig.~\ref{fig:Figure4}(e)--(f), under the standard Step function, the proportion of zeros is approximately 50\%. In contrast, VA-Step allows control over the ratio of activations mapped to $0$ versus $+1$ by adjusting the threshold $\theta$, resulting in a more skewed activation distribution. Specifically, introducing the activation threshold $\theta$ shifts the binarization decision boundary from $0$ in the standard binarization function to $\theta$. During training, this threshold drives more activations toward $0$, increasing the sparsity of the binary activation outputs. Fig.~\ref{fig:Figure5} illustrates the effects of different $\theta$ values on the activation distribution. As $\theta$ increases, the proportion of zeros in the activation outputs gradually increases.

To address the non-differentiability of the VA-Step binarization function during backpropagation, we employ a Sigmoid-based straight-through estimator (STE) for gradient approximation. Specifically, given $z=a_r-\theta$, the soft surrogate function is defined as
\begin{equation}
\sigma(z)=\frac{1}{1+e^{-tz}},
\label{eq:sigmoid_proxy}
\end{equation}
where $t$ is a temperature coefficient that controls the steepness of the Sigmoid approximation to the hard step function. By applying the chain rule, the gradient of the loss function $L$ with respect to $a_r$ is approximated as:
\begin{equation}
\frac{\partial L}{\partial a_r} = \frac{\partial L}{\partial \sigma(z)} \cdot \frac{\partial \sigma(z)}{\partial z} \cdot \frac{\partial z}{\partial a_r} = \frac{\partial L}{\partial \sigma(z)} \cdot t\sigma(z)\left(1-\sigma(z)\right).
\label{eq:sigmoid_proxy_grad}
\end{equation}
This surrogate function provides smooth, nonzero gradients near the threshold, thereby alleviating the gradient discontinuity caused by hard binarization and improving the stability of gradient propagation and parameter optimization during training.

\subsubsection{Voltage-Aware Training Algorithm for ReLU Functions}

As shown in Fig.~\ref{fig:Figure4}(c), in neural networks employing the ReLU activation function, negative pre-activations are truncated to zero. Therefore, ReLU inherently introduces a certain degree of activation sparsity. To further increase the proportion of zero activations, as illustrated in Fig.~\ref{fig:Figure4}(d), we introduce a preset threshold $\theta$ into the ReLU function, constructing the voltage-aware ReLU (VA-ReLU). By suppressing positive pre-activations with small magnitudes, this method extends the zero output region from $(-\infty, 0]$ to $(-\infty, \theta]$, enhancing activation sparsity while preserving nonlinear expressiveness. Given a real-valued pre-activation $a_r$, VA-ReLU is defined as
\begin{equation}
a = \max(0, a_r - \theta) =
\begin{cases}
a_r - \theta, & \text{if } a_r > \theta, \\
0, & \text{otherwise},
\end{cases}
\label{eq:relu_vat}
\end{equation}
where $a$ denotes the activation output and $\theta$ represents the preset threshold used to regulate activation sparsity.

As shown in Figs.~\ref{fig:Figure4}(g)--(h), under the standard ReLU function, the proportion of zero activations is approximately 50\%. In contrast, VA-ReLU reshapes the activation distribution by introducing a preset threshold $\theta$. Specifically, when $\theta=0$, VA-ReLU is equivalent to standard ReLU. When $\theta>0$, the activation decision boundary of VA-ReLU shifts from $0$ in standard ReLU to $\theta$. As training progresses, this threshold truncates pre-activations below $\theta$ to zero, leading to a significant increase in activation sparsity within the network.

Since VA-ReLU remains a piecewise linear function in form, its backpropagation can be computed using the piecewise derivative of ReLU. Specifically, for a given pre-activation $a_r$, the derivative of the activation output $a$ with respect to $a_r$ is
\begin{equation}
\frac{\partial a}{\partial a_r}
=
\begin{cases}
1, & a_r > \theta,\\
0, & \text{otherwise}.
\end{cases}
\label{eq:relu_vat_grad_ar}
\end{equation}
By applying the chain rule, the gradient of the loss function $L$ with respect to the pre-activation $a_r$ is expressed as:
\begin{equation}
\frac{\partial L}{\partial a_r} = \frac{\partial L}{\partial a} \cdot \frac{\partial a}{\partial a_r}.
\end{equation}
Consequently, gradients propagate to the preceding layer only when the pre-activation exceeds the threshold $\theta$, whereas gradients corresponding to low-magnitude pre-activations suppressed to zero are truncated. At the boundary point $a_r=\theta$, we adopt the same subgradient convention as standard ReLU, setting the derivative to 0.

\subsection{Crossbar-Level Computation Skipping via Co-Zero Activation Reordering}

As shown in Fig.~\ref{fig:Figure5}, following optimization via VAT, the proportion of zero-valued activations increases significantly, exhibiting higher sparsity. As illustrated by Case 3, when $\theta$ exceeds a certain level, some crossbars receive all-zero input vectors. According to the analysis in Section~\ref{subsec:activation}, the MVM operations performed by these crossbars make no effective contribution to the final output and can therefore be directly skipped. However, despite the substantial increase in element-wise activation sparsity, zero activations remain highly scattered across the input dimensions of the activation matrix, as shown in Fig.~\ref{fig:Figure2}(a). Consequently, all-zero input vectors rarely occur naturally, making it difficult to satisfy the condition for computation skipping. To address this issue, this section proposes CAR. By reordering the activation indices of the VAT-trained model, CAR consolidates scattered zeros into dense local regions. This consolidation maximizes the opportunities for crossbar-level computation skipping.

\subsubsection{Clustering-Based Co-Zero Activation Reordering Algorithm}
\begin{figure}[t]
    \centering
    \includegraphics[width=1\linewidth]{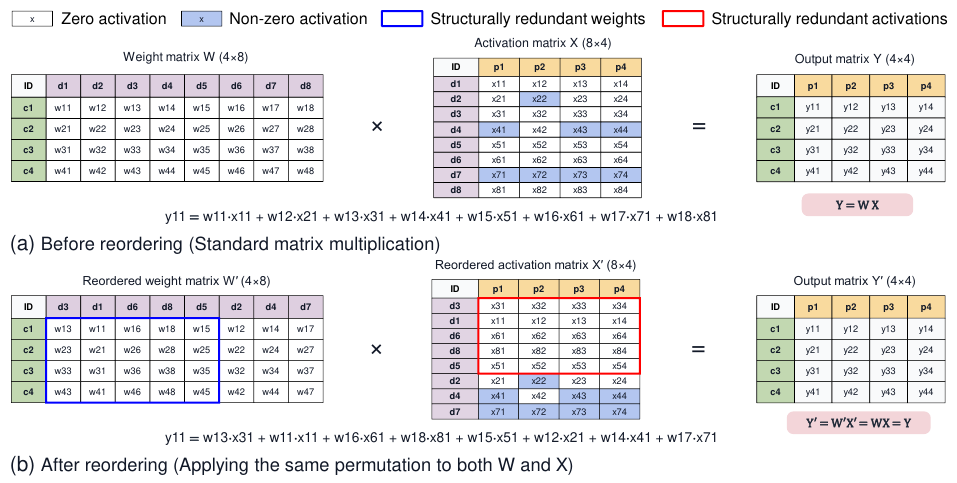}
    \vspace{-15pt}
    \caption{Example of computational equivalence in activation co-zero reordering. This reordering preserves the computational results while concentrating originally scattered zeros into more compact regions.}
    \Description{None}
    \vspace{-6pt}
\label{fig:Figure6}
\end{figure}

As illustrated in Fig.~\ref{fig:Figure6}, consistently reordering activations and their corresponding weight indices in matrix operations preserves the final computational results. Leveraging this mathematical equivalence, we propose a clustering-based co-zero activation reordering algorithm. Specifically, given the unrolled activation matrix $X_{\mathrm{col}} \in \mathbb{R}^{D \times L}$. Here, $D$ denotes the flattened input dimension of the convolution kernel, and $L$ represents the number of unrolled sliding windows. We first define a zero-indicator matrix as follows:
\begin{equation}
Z_{d,l} =
\begin{cases}
1, & X_{\mathrm{col}}(d,l)=0, \\
0, & \text{otherwise}.
\end{cases}
\label{eq:zero_indicator}
\end{equation}
where $Z_{d,l}=1$ indicates that the activation value corresponding to the $d$-th input dimension in the $l$-th sliding window is zero.

Based on this zero-indicator matrix, the zero rate of the $d$-th input dimension is defined as
\begin{equation}
r_d = \frac{1}{L}\sum_{l=1}^{L} Z_{d,l}.
\label{eq:zero_rate}
\end{equation}
This metric represents the proportion of sliding windows in which the $d$-th input dimension is zero, serving as a measure of sparsity for that dimension.

Furthermore, to characterize the co-zero relationship between different input dimensions, the co-zero affinity between the $i$-th and $j$-th input dimensions is defined as
\begin{equation}
A_{i,j} = \frac{1}{L}\sum_{l=1}^{L} Z_{i,l}Z_{j,l}.
\label{eq:cozero_affinity}
\end{equation}
Since $Z_{i,l}Z_{j,l}=1$ holds only when both the $i$-th and $j$-th input dimensions are zero in the $l$-th sliding window, $A_{i,j}$ represents the empirical co-zero probability, i.e., the proportion of sliding windows where both dimensions are simultaneously zero. A larger $A_{i,j}$ indicates a higher co-zero correlation between these two input dimensions, making them more suitable to be placed adjacently during reordering to maximize the probability of the entire input vector being zero.

Based on the co-zero affinity matrix, we employ a greedy strategy to generate a permutation of the input dimensions. Specifically, the strategy sequentially constructs input dimension blocks according to the row capacity of a ReRAM crossbar, where each block corresponds to an input vector mapped onto a single crossbar. For each block, the unassigned input dimension with the highest zero rate is first selected as the seed dimension. At each subsequent iteration, the candidate dimension with the highest composite score is added to the current block. For a candidate dimension $d$, its composite score with respect to the current block $\mathcal{B}$ is defined as
\begin{equation}
s(d|\mathcal{B}) =
\frac{1}{|\mathcal{B}|}\sum_{i\in \mathcal{B}} A_{d,i} + \alpha r_d,
\label{eq:cozero_score}
\end{equation}
where $\alpha$ is a weighting factor that controls the relative contribution of the candidate dimension's zero rate to the composite score. This selection process continues until the number of dimensions in the current block reaches the row capacity of the crossbar, after which the next block is constructed in the same manner. Finally, the dimension indices are concatenated in the order in which the blocks are constructed to obtain the input-dimension permutation $\pi$. Here, $\pi = [\pi_1, \pi_2, \dots, \pi_D]$ is an index sequence of length $D$ that specifies the reordered arrangement of the original input dimensions. Each element $\pi_k$ denotes the original dimension index placed at the $k$-th position in the reordered sequence. The permutation $\pi$ is then consistently applied to the unfolded activation matrix and its corresponding weight matrix. The complete procedure is presented in Algorithm~\ref{alg:cozero_reorder}.

\begin{algorithm}[ht]
\caption{Clustering-Based Co-Zero Activation Reordering Algorithm}
\label{alg:cozero_reorder}
\begin{algorithmic}[1]
\Require Unfolded activation matrix $\mathbf{X}_{\mathrm{col}} \in \mathbb{R}^{D \times L}$, crossbar size $S$, sparsity weight $\alpha$, maximum samples $L_{\max}$
\Ensure Permutation vector $\pi$

\State Optionally sample $L_s = \min(L, L_{\max})$ columns from $\mathbf{X}_{\mathrm{col}}$
\State Construct zero-indicator matrix $\mathbf{Z} \in \{0,1\}^{D \times L_s}$ from the sampled activation matrix
\State Compute zero rate $r_d \leftarrow \frac{1}{L_s}\sum_{l=1}^{L_s} Z_{d,l}$ for each dimension $d$
\State Compute co-zero affinity matrix $\mathbf{A}$, where $A_{i,j} = \frac{1}{L_s}\sum_{l=1}^{L_s} Z_{i,l}Z_{j,l}$
\State Initialize unused dimension set $\mathcal{U} \leftarrow \{1,2,\ldots,D\}$ and permutation list $\pi \leftarrow \emptyset$

\While{$\mathcal{U} \neq \emptyset$}
    \State Select seed dimension $d^{*} \leftarrow \arg\max_{d\in\mathcal{U}} r_d$
    \State Initialize current block $\mathcal{B} \leftarrow \{d^{*}\}$ and update $\mathcal{U} \leftarrow \mathcal{U} \setminus \{d^{*}\}$
    \State Initialize running co-zero sum vector $\mathbf{v} \leftarrow \mathbf{A}_{d^{*}, :}$
    
    \While{$|\mathcal{B}| < S$ \textbf{and} $\mathcal{U} \neq \emptyset$}
        \State Compute candidate clustering scores:
        \Statex \hspace*{4em}
        $s_d \leftarrow \frac{1}{|\mathcal{B}|} v_d + \alpha r_d,\quad \forall d \in \mathcal{U}$
        
        \State Select next dimension $d^{*} \leftarrow \arg\max_{d\in\mathcal{U}} s_d$
        \State Update $\mathcal{B} \leftarrow \mathcal{B} \cup \{d^{*}\}$ and $\mathcal{U} \leftarrow \mathcal{U} \setminus \{d^{*}\}$
        \State Update running vector $\mathbf{v} \leftarrow \mathbf{v} + \mathbf{A}_{d^{*}, :}$
    \EndWhile
    \State Append dimensions in $\mathcal{B}$ to $\pi$
\EndWhile

\State Apply $\pi$ to both activation and weight dimensions before crossbar partitioning
\State \Return $\pi$
\end{algorithmic}
\end{algorithm}

\subsubsection{Verification of Computational Equivalence}
To verify the computational equivalence of matrix multiplication before and after reordering, let $P_{\pi}$ denote the permutation matrix corresponding to the permutation sequence $\pi$. The reordered activation and weight matrices are then expressed as:
\begin{equation}
X_{\mathrm{col}}^{\prime} = P_{\pi} X_{\mathrm{col}}, \qquad
W_{\mathrm{col}}^{\prime} = W_{\mathrm{col}} P_{\pi}^\top ,
\label{eq:cozero_reorder}
\end{equation}
where left-multiplication by $P_{\pi}$ reorders the input dimensions of the activation matrix. Right-multiplication by $P_{\pi}^{\top}$ applies the same index permutation to the corresponding dimensions of the weight matrix. Since the permutation matrix satisfies $P_{\pi}^{\top}P_{\pi}=I_D$ (where $I_D$ is the $D \times D$ identity matrix), the result of the reordered matrix multiplication is given by:
\begin{equation}
W_{\mathrm{col}}^{\prime} X_{\mathrm{col}}^{\prime} =
W_{\mathrm{col}} P_{\pi}^\top P_{\pi} X_{\mathrm{col}}
= W_{\mathrm{col}} X_{\mathrm{col}}.
\label{eq:reorder_equivalence}
\end{equation}
Therefore, the CAR scheme merely adjusts the mapping order of activations and weights within the ReRAM crossbar without altering the original matrix multiplication results. As illustrated in Fig.~\ref{fig:Figure6}, this reordering places input dimensions with high co-zero affinity adjacent to one another, creating highly concentrated zero-valued regions within the matrix and increasing the probability of all-zero input vectors.

\subsubsection{Computation Skipping Based on Activation Co-Zero Reordering}

Neural network activations are generated dynamically from input data, and their zero patterns cannot be known in advance. Analyzing co-zero relationships among activation dimensions and searching for an optimal reordering at each layer during online inference on a ReRAM accelerator would incur substantial overhead in statistics collection, computation, and scheduling, substantially offsetting the energy-efficiency benefits of computation skipping. To address this issue, CAR combines offline profiling with online inference. Before model deployment, training samples are used to determine fixed permutation indices for activation dimensions. During online inference, these indices are directly applied to reorder activations, and the reordered activation matrix is partitioned into input blocks. The computation associated with each block is skipped if the current block is all zero.

During the offline profiling phase, the complete training set is fed into the VAT-pretrained model operating in gradient-free evaluation mode. For the activation matrix produced at each layer, Algorithm~\ref{alg:cozero_reorder} analyzes the co-zero correlations across distinct activation dimensions to generate a permutation index $\pi$. Subsequently, the weight matrix undergoes physical reordering and partitioning according to the identical permutation index $\pi$, with the partitioning scheme illustrated in Fig.~\ref{fig:Figure2}(b). Because index determination, weight reordering, and weight programming into crossbars are performed offline prior to deployment, this phase introduces no runtime overhead.

Notably, selecting the training set as the profiling benchmark is motivated by two key factors. First, data distribution homogeneity: the training and test sets typically share identical or closely matched data distributions, ensuring high transferability of the co-zero correlations and permutation indices derived from the training set to unseen test data. Second, generalization robustness: the VAT-optimized model possesses strong intrinsic generalization capabilities, which effectively absorb and smooth out minor distribution discrepancies between profiling samples and inference inputs, guaranteeing that the optimal permutation index derived from the training set reliably preserves substantial all-zero sparsity clustering on the test set.

During online inference, before executing the next layer, VARA remaps the read addresses in the activation buffer according to the permutation indices $\pi$, allowing activations to be read in the permuted order. This process changes only the buffer access order and requires no explicit data movement, thereby incurring very low runtime overhead. The reordered activation matrix is then partitioned according to the crossbar input granularity and distributed to the corresponding ReRAM crossbars. For all-zero input blocks, the system bypasses the corresponding input driving, crossbar reads, and ADC conversions; blocks containing nonzero activations undergo normal crossbar computation and analog-to-digital conversion. By consistently aligning offline weight reordering with online activation address remapping and applying runtime gating based on the all-zero condition, CAR enables crossbar-level computation skipping while preserving exact computational equivalence, eliminating redundant crossbar computation and peripheral readout overhead.

\section{Experimental Evaluation}
\label{sec:section_4}

\subsection{Experimental Setup}
\subsubsection{Model Training}
In this work, we select three representative networks—VGG11, ResNet18, and ResNet34—and train and evaluate them on the CIFAR10, CIFAR100, and Tiny ImageNet datasets, respectively. All experiments are implemented using the PyTorch framework. During training, stochastic gradient descent (SGD) with momentum is employed as the optimizer, combined with a cosine annealing schedule to dynamically adjust the learning rate. Model weights are uniformly quantized to 8~bits. To increase the proportion of zero activations in each network, VA-Step or VA-ReLU is adopted as the activation function, and training is conducted under various threshold settings. The key training hyperparameters are summarized in Table~\ref{tab:training_params}.

\setlength{\tabcolsep}{18pt}
\renewcommand{\arraystretch}{1.2}
\begin{table}[t]
\centering
\small
\caption{Hyperparameter Settings for Model Training}
\vspace{-4pt}
\begin{tabular}{ccc}
\toprule
\textbf{Category} & \textbf{Hyperparameter} & \textbf{Value} \\
\hline
\multirow{2}{*}{\makecell{Optimization \\ \& regularization}} & Momentum & 0.9 \\
\cline{2-3}
 & Weight decay & \(1 \times 10^{-4}\) \\
\hline
\multirow{3}{*}{Training settings} & Epochs & 300 \\
\cline{2-3}
 & Batch size & 128 \\
\cline{2-3}
 & Initial learning rate & 0.1 \\
\bottomrule
\end{tabular}
\label{tab:training_params}
\end{table}

\subsubsection{Hardware Setup}
To evaluate the hardware performance of the VARA architecture, we developed a customized simulator based on open-source simulators~\cite{6218223,chen2017neurosim,10058114} for simulation analysis. The simulation employs an Ag:Si ReRAM device model at the 22~nm technology node~\cite{jo2010nanoscale} and adopts a one-transistor–one-resistor (1T1R) cell structure to construct ReRAM crossbars. During model deployment, the sparse model obtained from VAT training serves as the baseline; the CAR scheme is then applied to reorder activation dimensions and their corresponding weights. Subsequently, matrix partitioning and mapping are performed according to crossbar size constraints, enabling crossbar-level computation skipping during inference. The main hardware parameters are summarized in Table~\ref{tab:hardware_params}.

\setlength{\tabcolsep}{14.3pt}
\renewcommand{\arraystretch}{1.2}
\begin{table}[t]
\centering
\small
\caption{Hardware Simulation Parameters}
\vspace{-4pt}
\begin{tabular}{ccc}
\toprule
\textbf{Component} & \textbf{Parameter} & \textbf{Value} \\
\hline
\multirow{2}{*}{ReRAM device}  
 & LRS/HRS & 0.1\,M$\Omega$/1.7\,M$\Omega$ \\
\cline{2-3}
 & Read voltage & 0.5\,V  \\
\hline
\multirow{2}{*}{Crossbar array} 
 & Technology node & 22\,nm \\
\cline{2-3}
 & Crossbar size & 64$\times$64 \\
\hline
ADC & Resolution & 6~bit \\
\bottomrule
\end{tabular}
\label{tab:hardware_params}
\end{table}

\subsection{Impact of VAT on Model Accuracy and Activation Sparsity}
\begin{figure*}[!t]
    \centering
    \includegraphics[width=1\linewidth]{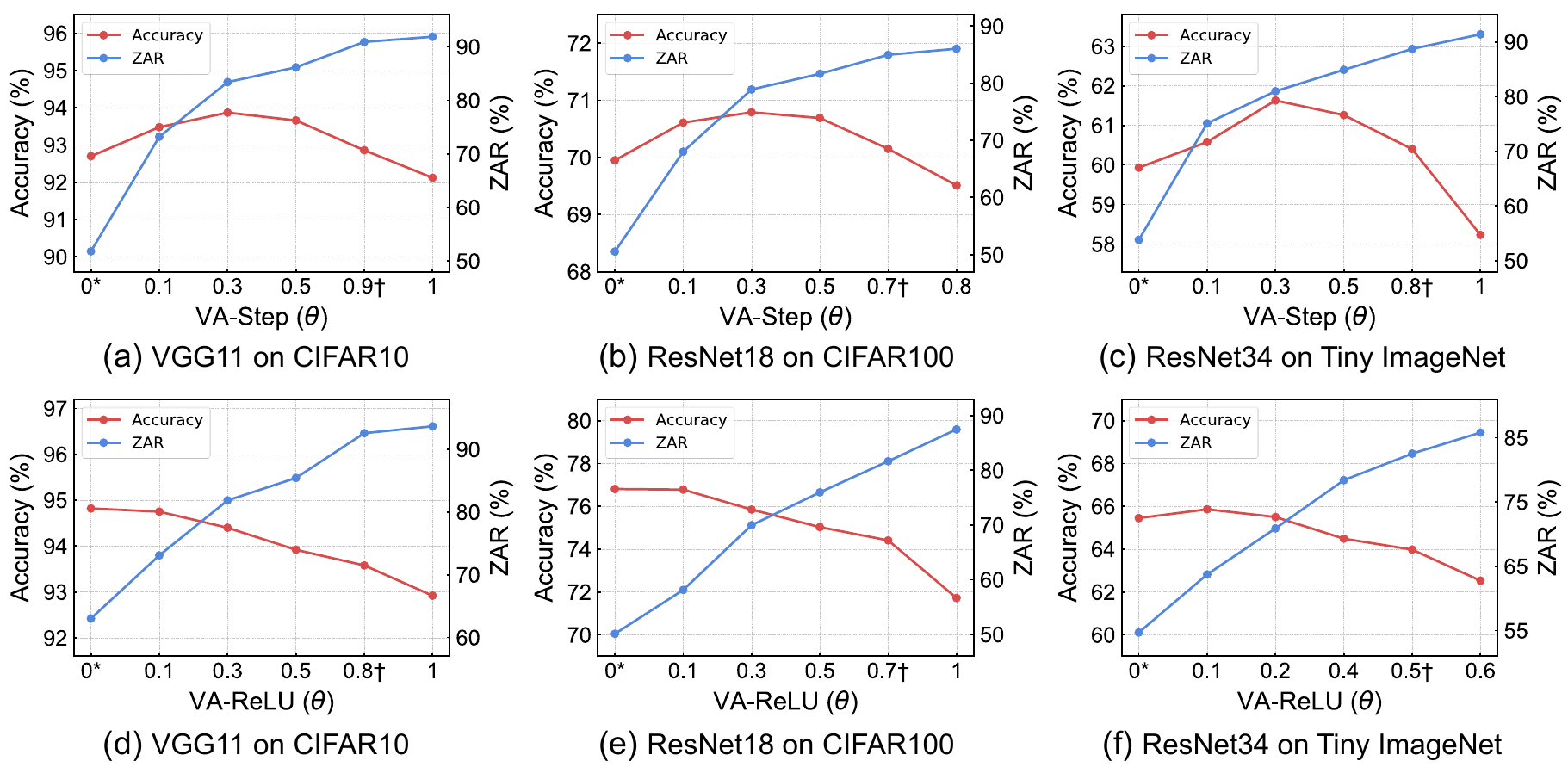}
    \vspace{-16pt}
    \caption{Impact of activation threshold $\theta$ on model accuracy and ZAR. Here, $\ast$ denotes the baseline, and $\dagger$ represents the optimal threshold.}
    \Description{None}
    \vspace{-12pt}
    \label{fig:Figure7}
\end{figure*}

This section evaluates the impact of the VAT algorithm on model classification accuracy and ZAR. The conventional Step function at $\theta=0$ and the standard ReLU function serve as baselines for VA-Step and VA-ReLU, respectively. Under identical training settings, we analyze the effect of the activation threshold $\theta$ on both model accuracy and activation sparsity. The results are presented in Fig.~\ref{fig:Figure7}, where $\ast$ denotes the baseline threshold and $\dagger$ indicates the optimal threshold selected by jointly considering model accuracy and activation sparsity.

\paragraph{Evaluation Results of VA-Step}
As shown in Fig.~\ref{fig:Figure7}(a)--(c), the ZARs of all three models increase steadily as the activation threshold $\theta$ increases, while model accuracy generally exhibits an initial increase followed by a decline. Specifically, the accuracies of VGG11, ResNet18, and ResNet34 peak at approximately $\theta=0.3$ and exceed their corresponding baselines. This indicates that moderately suppressing redundant low-magnitude activations provides a regularization effect that improves model generalization. As the threshold increases further, the ZAR continues to rise, whereas model accuracy begins to decline. Nevertheless, VA-Step preserves accuracy well over a relatively wide range of thresholds. Accordingly, we select $\theta=0.9$, $\theta=0.7$, and $\theta=0.8$ as the optimal thresholds for VGG11, ResNet18, and ResNet34, respectively. Under these optimal settings, the three models achieve ZARs of 90.89\%, 84.96\%, and 88.73\%, respectively, while improving accuracy over their corresponding baselines by 0.16\%, 0.20\%, and 0.47\%, respectively.

\paragraph{Evaluation Results of VA-ReLU}
As illustrated in Fig.~\ref{fig:Figure7}(d)--(f), the ZAR of VA-ReLU also increases steadily with the activation threshold $\theta$, but its model accuracy is more sensitive to threshold variations. For VGG11 and ResNet18, accuracy generally declines as $\theta$ increases; for ResNet34, a marginal improvement occurs only at a small threshold of $\theta=0.1$, followed by a continuous decrease. This phenomenon primarily stems from the truncation effect of VA-ReLU on activations. As $\theta$ increases, more sub-threshold activations are zeroed out, including some that still carry meaningful feature information. The loss of this information progressively degrades model accuracy. Accordingly, we select $\theta=0.8$, $\theta=0.7$, and $\theta=0.5$ as the optimal thresholds for the three models, respectively. Under these optimal settings, the models achieve ZARs of 92.58\%, 81.66\%, and 82.53\%, while exhibiting accuracy drops of approximately 1.24\%, 2.39\%, and 1.46\% compared to their corresponding baselines, respectively.

Across the six experiments, under the selected thresholds, VAT increases the average ZAR across models to 86.89\% while limiting the average accuracy loss to approximately 0.71\%. Overall, both VA-Step and VA-ReLU can significantly improve ZAR by adjusting the activation threshold, yet they exhibit distinct accuracy-sparsity trade-offs. VA-Step maintains a wider accuracy retention window and can preserve or even slightly improve model accuracy at high ZAR levels; in contrast, VA-ReLU is more sensitive to threshold variations, with its accuracy declining more markedly as the threshold increases.
Since the system performance of VARA depends not only on model accuracy but also on the ZAR, the proportion of all-zero input vectors, and crossbar-level computation skipping opportunities, we do not select the threshold yielding peak accuracy. Instead, we adopt the accuracy-sparsity trade-off points marked with $\dagger$ in the figure as the optimal thresholds. These settings maximize ZAR within an acceptable accuracy loss margin and provide sufficient sparsity for subsequent structured sparsity exploitation and crossbar-level computation skipping.

\begin{figure*}[!t]
    \centering
    \includegraphics[width=1.0\linewidth]{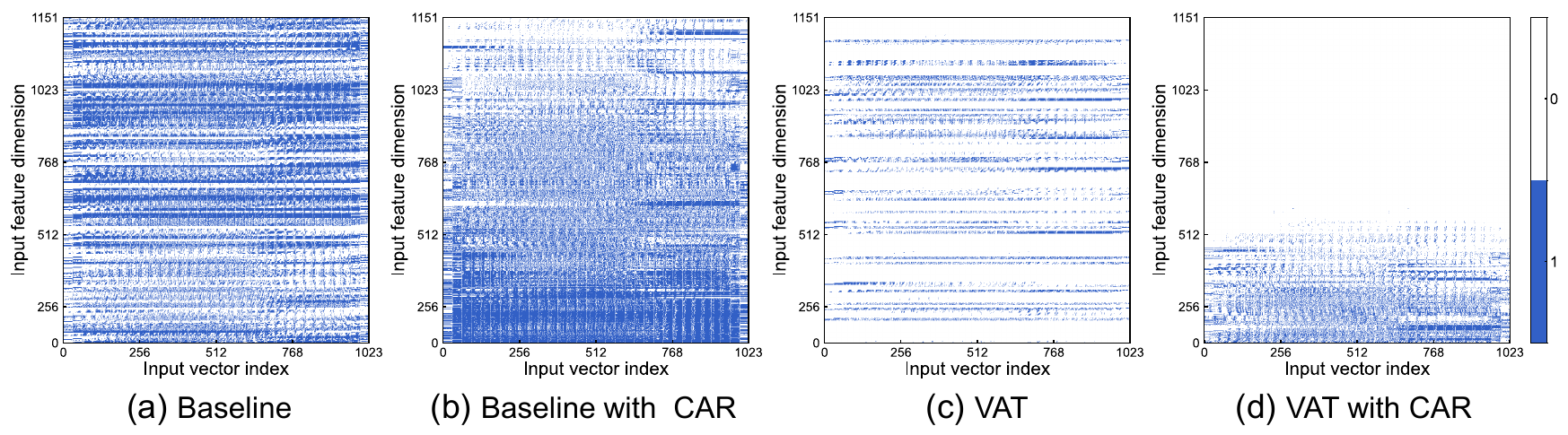}
    \vspace{-15pt}
    \caption{Visual comparison of activation distributions before and after CAR reordering under different thresholds. (a) and (b) show the activation distributions before and after reordering at $\theta=0$, respectively; (c) and (d) show those at $\theta=1.0$, respectively.}
    \Description{None}
    \vspace{-12pt}
    \label{fig:Figure8}
\end{figure*}

\subsection{Evaluation of Structured Reshaping Effects on Activation Distributions}
Taking the first layer of each model as an example, this section evaluates the impact of VAT and CAR on activation sparsity and crossbar-level computation skipping. We first visualize the reshaping effects of both methods on activation distributions. 
Subsequently, we analyze the influence of crossbar size and CAR on the crossbar-level computation skipping ratio under three experimental configurations to validate the effectiveness of transforming element-wise activation sparsity into exploitable structured sparsity.

\subsubsection{Visual Analysis of the Synergistic Reshaping of Activation Distributions by VAT and CAR}
Fig.~\ref{fig:Figure8} illustrates the activation distributions of the first layer of VGG11 with the VA-Step activation function for the first input batch under different optimization strategies, where white and blue dots represent zero and nonzero activations, respectively. Fig.~\ref{fig:Figure8}(a) shows the baseline activation distribution without VAT or CAR. Zero and nonzero activations are interspersed across feature dimensions, exhibiting a pronounced unstructured pattern that hinders the formation of contiguous all-zero regions aligned with the input partitions of ReRAM crossbars. Fig.~\ref{fig:Figure8}(b) presents the activation distribution after applying CAR reordering alone. By reordering activations based on co-zero correlations among input dimensions, CAR enhances the local clustering of zero activations along the input dimension. However, due to the relatively limited overall ZAR of the baseline model, the existing zero activations are insufficient to support large-scale structured clustering. Consequently, CAR alone fails to generate extensive contiguous all-zero regions.

Fig.~\ref{fig:Figure8}(c) shows the activation distribution after applying VAT alone. By introducing a predefined activation threshold, VAT maps a large number of low-magnitude activations to zero, significantly increasing the overall sparsity of the activation matrix and producing noticeable clusters of zeros. However, some nonzero activations remain scattered across the input dimensions, causing many input partitions to contain a small number of nonzero elements. This dispersion limits the formation of all-zero input vectors. Fig.~\ref{fig:Figure8}(d) shows the activation distribution obtained by combining VAT and CAR. Building on the high element-wise sparsity induced by VAT, CAR further reorders the input dimensions according to their co-zero correlations. This reordering concentrates nonzero activations within a lower-indexed subset of dimensions and creates large, contiguous all-zero regions across the remaining dimensions. When the activation matrix is partitioned and mapped according to the row capacity of ReRAM crossbars, this structured sparsity pattern significantly increases the probability of producing all-zero input vectors, creating more opportunities for crossbar-level computation skipping.

\begin{figure*}[!t]
    \centering
    \includegraphics[width=1.0\linewidth]{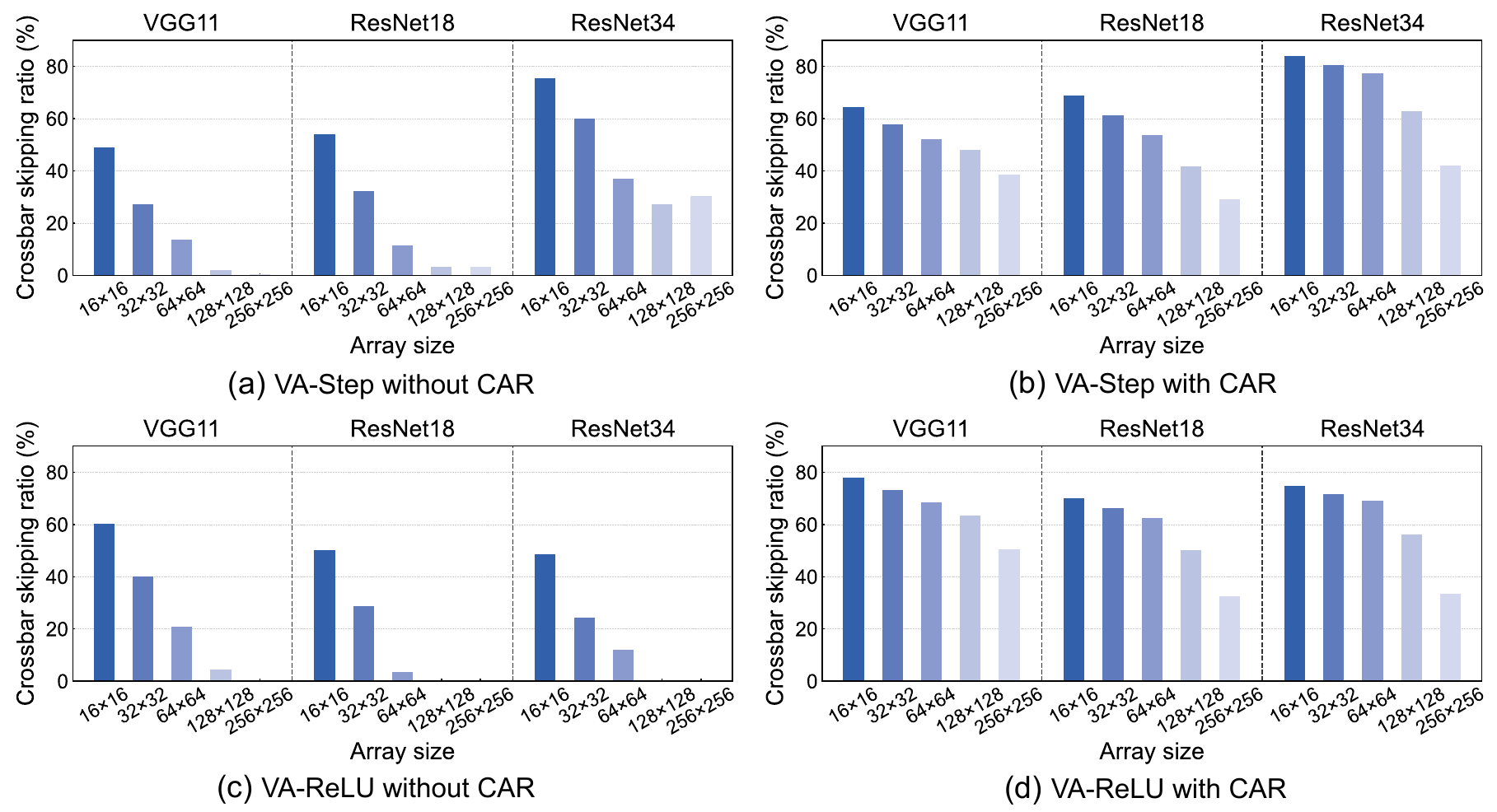}
    \vspace{-20pt}
    \caption{Crossbar-level computation skipping ratios before and after CAR reordering as a function of crossbar size at the optimal activation threshold. (a) and (b) show the results for VA-Step before and after reordering, respectively; (c) and (d) show the corresponding results for VA-ReLU.}
    \Description{None}
    \vspace{-12pt}
    \label{fig:Figure9}
\end{figure*}

\subsubsection{Impact of Array Size and CAR on crossbar-Level Computation Skipping Ratio}

During offline model deployment, CAR reorders the activation dimensions to consolidate scattered zero activations into contiguous all-zero regions, creating more opportunities for crossbar-level computation skipping. To quantify this benefit, we define the crossbar-level computation skipping ratio as the fraction of all-zero input vectors among all input vectors, since each all-zero input vector corresponds to one skippable crossbar computation. A higher ratio indicates that activation sparsity can be more effectively translated into crossbar-level computation skipping. Fig.~\ref{fig:Figure9} presents the crossbar-level computation skipping ratios with and without CAR across different crossbar sizes at the optimal activation threshold.

As shown in Fig.~\ref{fig:Figure9}(a) and (c), without CAR, the crossbar-level computation skipping ratios of all three models decrease rapidly as the crossbar size increases from 16$\times$16 to 256$\times$256, despite the high ZAR achieved by VAT. For VA-Step, the computation skipping ratios of VGG11 and ResNet18 approach zero at larger crossbar sizes. Although ResNet34 retains some computation skipping opportunities, its skipping ratio also decreases significantly. VA-ReLU exhibits a similar pattern, with the computation skipping ratios of all three models approaching zero at crossbar sizes of 128$\times$128 and 256$\times$256. 
These results indicate that, in the absence of structured reordering, scattered nonzero activations make all-zero input vectors extremely difficult to form. Consequently, even a high ZAR cannot be effectively translated into a high crossbar-level computation skipping ratio.

As shown in Fig.~\ref{fig:Figure9}(b) and (d), applying CAR significantly improves the crossbar-level computation skipping ratios for all three models. For VA-Step, even as the crossbar size increases to 256$\times$256, the computation skipping ratios of VGG11, ResNet18, and ResNet34 still reach approximately 38.4\%, 29.0\%, and 42.1\%, respectively. For VA-ReLU, at a crossbar size of 256$\times$256, the computation skipping ratios of the three models remain at approximately 50.5\%, 32.7\%, and 33.3\%, respectively. Compared to the case without CAR, the improvement introduced by CAR is particularly pronounced at larger crossbar sizes. This indicates that CAR mitigates the constraint imposed by increasing input vector length on the formation of all-zero input vectors, thereby translating a greater portion of activation sparsity into crossbar-level computation skipping.

This result can be explained from two perspectives: input-partitioning granularity and activation distribution. When a crossbar has fewer rows, each input vector contains fewer activation elements. Consequently, fewer elements need to be simultaneously zero to form an all-zero input vector, making crossbar-level computation skipping more likely. As the crossbar size increases, the input-vector length also increases, requiring more activation elements to be simultaneously zero and thus making all-zero input vectors more difficult to form. CAR reorders activation dimensions with strong co-zero correlations to concentrate zero activations locally, thereby increasing the probability of forming all-zero input vectors. Although the crossbar-level computation skipping ratio decreases with increasing crossbar size, CAR substantially mitigates this degradation and preserves considerable skipping opportunities at larger crossbar sizes. Based on this crossbar-size sensitivity analysis, a crossbar size of $64\times64$ provides a moderate input partitioning granularity while maintaining a considerable crossbar-level computation skipping ratio across different models. Therefore, it is adopted as the default configuration for the subsequent hardware performance evaluation.

\begin{figure*}[!t]
    \centering
    \includegraphics[width=1.0\linewidth]{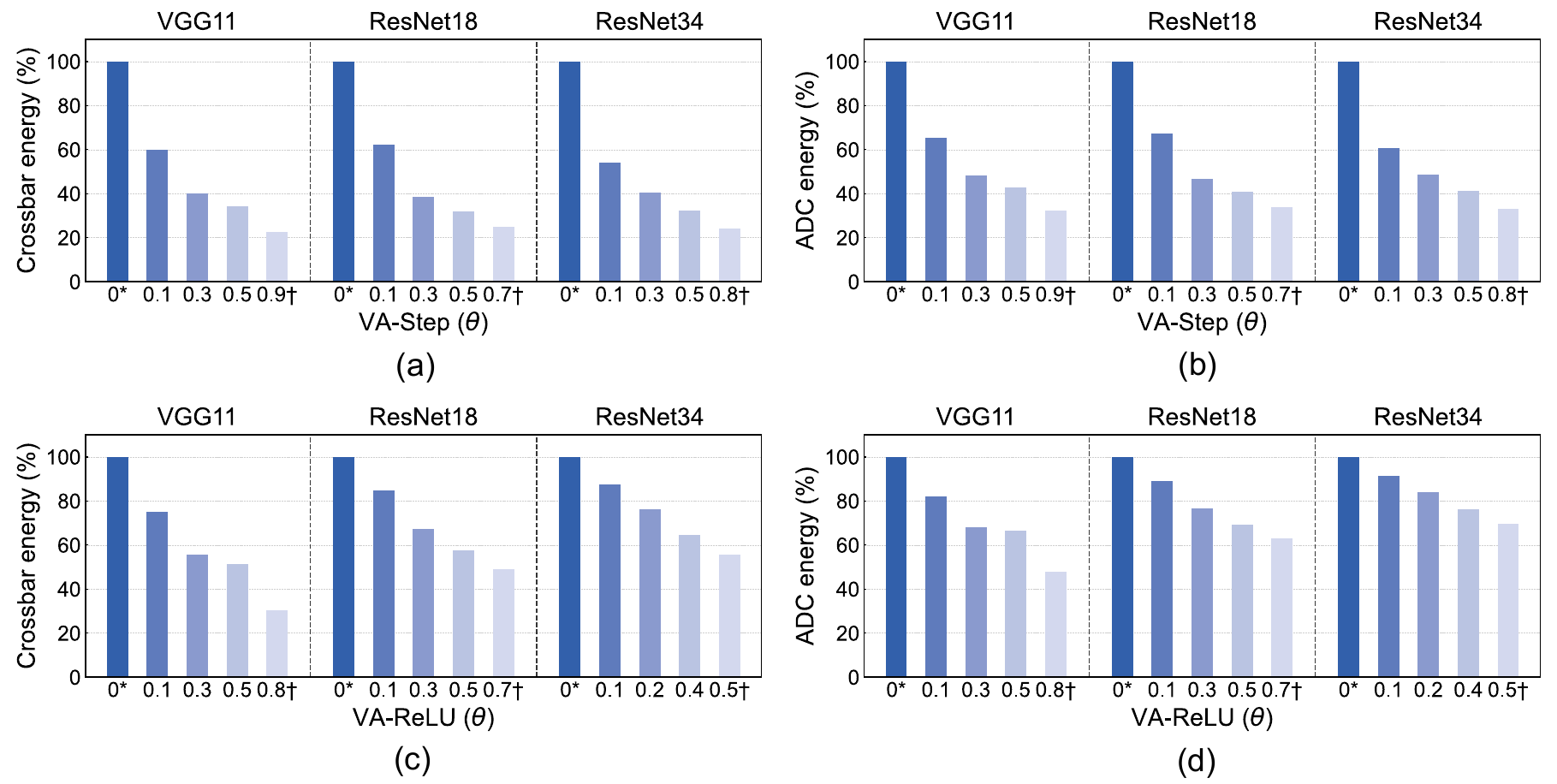}
    \vspace{-18pt}
    \caption{Impact of threshold $\theta$ on crossbar and ADC energy consumption in VA-Step and VA-ReLU.}
    \Description{None}
    \vspace{-10pt}
    \label{fig:Figure10}
\end{figure*}

\subsection{Circuit-Level Hardware Performance Evaluation of the VARA Architecture with VAT Only}

As analyzed in Section~\ref{subsec:activation}, the numerical distribution of input activations directly determines the input voltages applied to ReRAM crossbars and the resulting bitline currents. These electrical characteristics, in turn, affect the dynamic energy consumption of both the crossbars and ADCs.
Accordingly, this section evaluates the crossbar and ADC energy consumption under VA-Step and VA-ReLU across different activation thresholds. The conventional Step function and standard ReLU at $\theta=0$ serve as baselines, with their respective crossbar and ADC energy consumptions normalized to $100\%$. The experimental results are presented in Fig.~\ref{fig:Figure10}.

\paragraph{Evaluation Results of VA-Step}
As shown in Fig.~\ref{fig:Figure10}(a) and (b), increasing the threshold leads to substantial reductions in both crossbar and ADC energy consumption across all three models when using VA-Step. At the optimal threshold, the crossbar energy consumption for VGG11, ResNet18, and ResNet34 is reduced by approximately 77.25\%, 74.18\%, and 75.76\% relative to the baseline, respectively, while ADC energy consumption decreases by approximately 67.71\%, 65.30\%, and 66.95\%, respectively. These gains primarily stem from the discrete $\{0,+1\}$ output characteristic of VA-Step. By mapping a large proportion of pre-activations to zero, VA-Step significantly increases the fraction of zero input voltages, thereby reducing effective ReRAM cell conduction and current flow within the crossbar. Concurrently, the lower accumulated bitline current reduces ADC conversion overhead, resulting in simultaneous energy savings for both the crossbar and the ADCs.

\paragraph{Evaluation Results of VA-ReLU}
As illustrated in Fig.~\ref{fig:Figure10}(c) and (d), VA-ReLU also achieves reductions in both crossbar and ADC energy consumption as the threshold increases. At the optimal threshold, the crossbar energy consumption for VGG11, ResNet18, and ResNet34 is reduced by approximately 69.52\%, 50.79\%, and 44.20\% relative to the baseline, respectively, while ADC energy consumption decreases by approximately 51.97\%, 37.05\%, and 30.45\%, respectively. Although VA-ReLU yields significant energy savings across all three models, its overall reduction is less pronounced than that of VA-Step. This difference arises because VA-ReLU truncates only sub-threshold activations to zero while preserving positive activations above the threshold. Consequently, the resulting activation sparsity and input voltage modulation are more gradual, leading to weaker suppression of cell conduction and bitline currents compared to VA-Step, leading to comparatively limited energy savings.

Across the six experiments, VAT reduces crossbar and ADC energy consumption by an average of 65.28\% and 53.24\%, respectively, demonstrating its ability to effectively translate algorithm-level activation sparsity into circuit-level energy savings. These savings arise because VAT increases the proportion of zero input voltages applied to ReRAM crossbars, suppressing the conduction currents of the corresponding ReRAM cells and further reducing the bitline current. This simultaneously lowers crossbar computation energy and ADC conversion energy. The experimental results also reveal that, compared to the ADC, the crossbar energy exhibits a more pronounced downward trend as activation sparsity increases. Furthermore, owing to its discrete ${0,+1}$ output representation, VA-Step achieves more significant and consistent circuit-level energy savings across different models than VA-ReLU.

\subsection{System-Level Performance Evaluation of the VARA Architecture under Synergistic VAT and CAR Optimization}

\begin{figure*}[!t]
    \centering
    \includegraphics[width=1.0\linewidth]{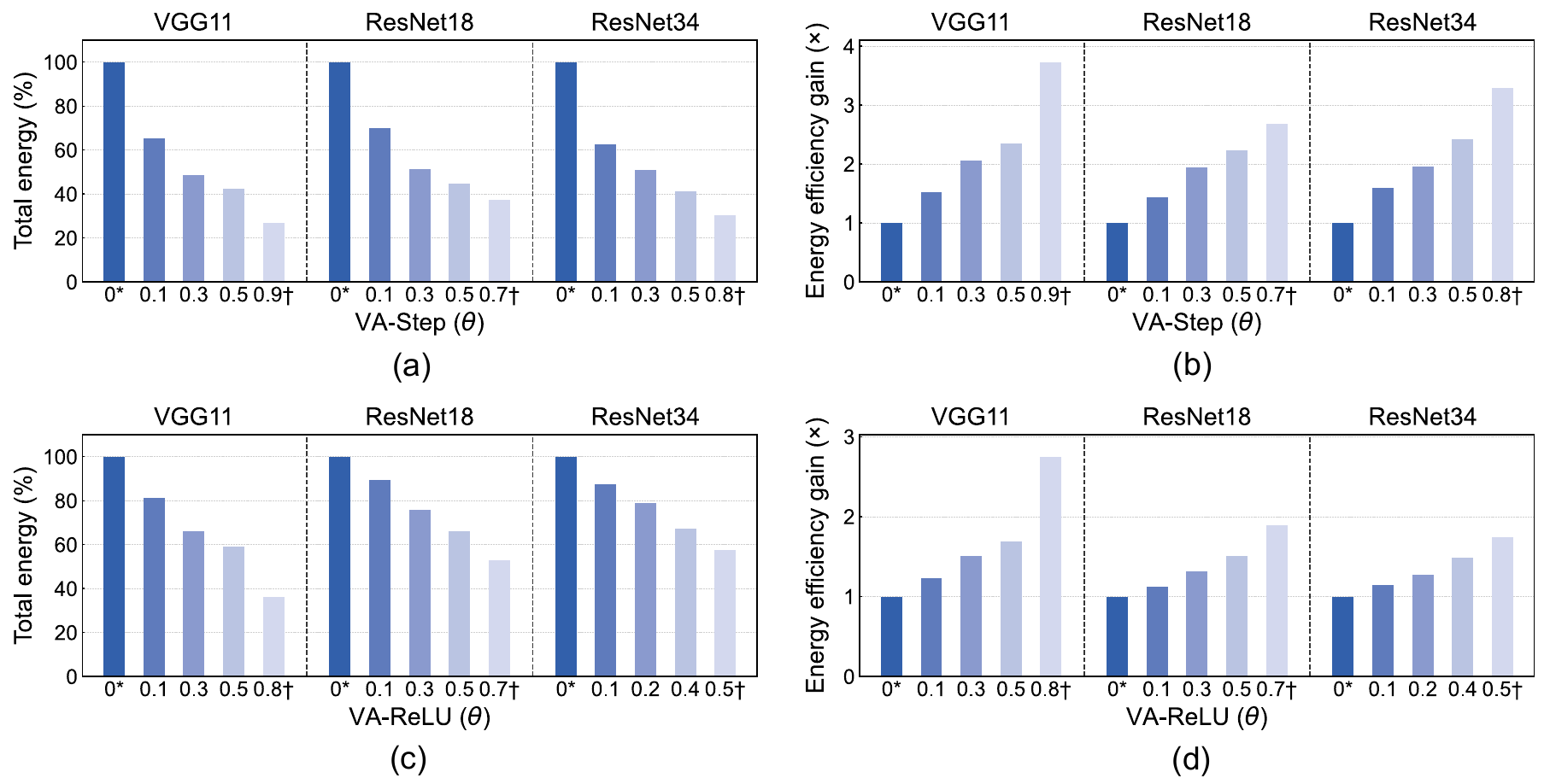}
    \vspace{-20pt}
    \caption{Impact of activation threshold $\theta$ on total system energy consumption and energy efficiency with CAR for VA-Step and VA-ReLU.}
    \Description{None}
    \vspace{-18pt}
    \label{fig:Figure11}
\end{figure*}

This section further evaluates the system-level performance of the VARA architecture leveraging the synergy between VAT and CAR. With CAR enabled and all other hardware configurations held constant, systems employing the conventional Step function and standard ReLU at $\theta=0$ serve as baselines for VA-Step and VA-ReLU, respectively. The total system energy consumption and energy efficiency under the baseline settings are normalized to 100\% and 1$\times$, respectively. The experimental results are presented in Fig.~\ref{fig:Figure11}.

\paragraph{Evaluation Results of VA-Step}
As shown in Fig.~\ref{fig:Figure11}(a) and (b), with increasing threshold, VA-Step exhibits a consistent trend of decreasing total system energy consumption and improving energy efficiency across all three models. At the optimal threshold, the total system energy consumption for VGG11, ResNet18, and ResNet34 is reduced by approximately 73.16\%, 62.80\%, and 69.57\% relative to the baseline, respectively, while the corresponding system energy efficiency improves to approximately 3.73$\times$, 2.69$\times$, and 3.29$\times$, respectively. These gains are primarily attributed to the discrete $\{0,+1\}$ output representation of VA-Step. As the threshold increases, more low-magnitude activations are mapped directly to zero, which not only reduces input voltage activity and bitline current in the ReRAM crossbar but also provides greater flexibility for CAR to exploit co-zero reordering opportunities. By grouping input dimensions with strong co-zero correlations into the same input partitions, CAR enables more crossbars to receive all-zero input vectors and allows the corresponding crossbar-level computations to be skipped. Consequently, building on the circuit-level energy savings, VA-Step achieves substantial additional improvements in system-level energy efficiency.

\paragraph{Evaluation Results of VA-ReLU}
As illustrated in Fig.~\ref{fig:Figure11}(c) and (d), VA-ReLU similarly exhibits a trend of decreasing total system energy consumption and increasing energy efficiency as the threshold rises. Under the optimal threshold setting, the normalized total system energy consumption for VGG11, ResNet18, and ResNet34 is reduced by approximately 63.68\%, 47.08\%, and 44.39\% relative to the baseline, respectively, while the corresponding system energy efficiency improves to approximately 2.75$\times$, 1.89$\times$, and 1.74$\times$, respectively. Compared to VA-Step, the system-level optimization achieved by VA-ReLU is relatively limited. This is because VA-ReLU zeros out only sub-threshold activations while preserving continuous positive activations above the threshold, resulting in more moderate improvements in both the ZAR and co-zero clustering. Consequently, CAR yields fewer all-zero input vectors, resulting in fewer crossbar and peripheral circuit operations that can be skipped. Nevertheless, the activation sparsity induced by VA-ReLU can still be effectively leveraged by CAR to yield substantial system-level energy and efficiency gains.

Across all six experimental configurations, the synergistic optimization of VAT and CAR reduces the average total system energy consumption by 60.12\% and improves the overall energy efficiency by an average of 2.68$\times$. These substantial gains demonstrate the high complementarity of the two methods. VAT increases the ZAR at the algorithmic level and directly reduces energy consumption in crossbars and ADCs. Meanwhile, CAR consistently reorders activation dimensions and their corresponding weight mappings to aggregate scattered element-wise sparsity into structured all-zero regions. These regions are aligned with the crossbar granularity and enable crossbar-level computation skipping. Together, they establish a complete optimization pipeline spanning activation sparsity enhancement, structured reordering, and hardware computation skipping, ultimately achieving significant reductions in total system energy consumption and effective improvements in overall energy efficiency.

\subsection{System-Level Energy Breakdown of the VARA Architecture}
\begin{figure*}[!t]
    \centering
    \includegraphics[width=0.65\linewidth]{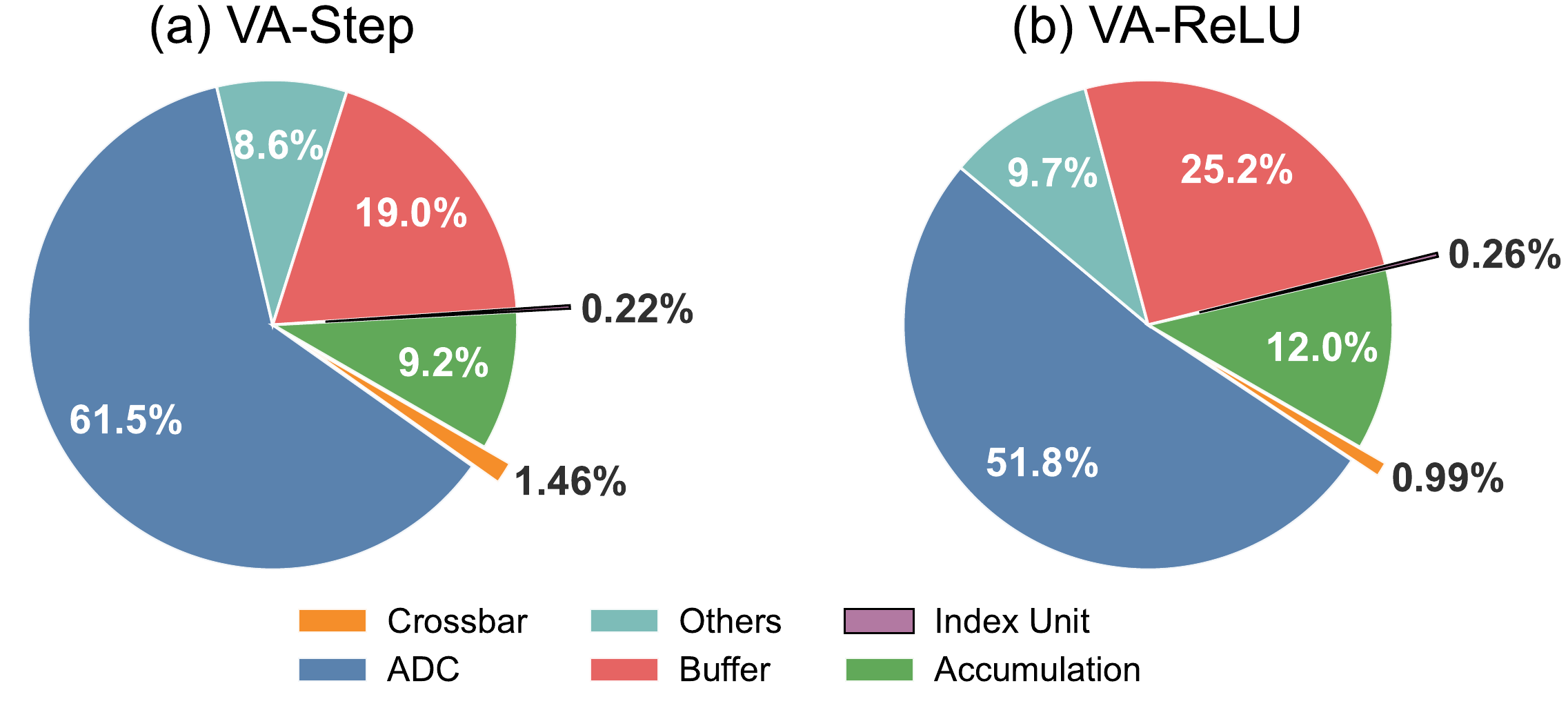}
    % \vspace{-15pt}
    \caption{System-level energy breakdown of VARA.}
    \Description{None}
    \vspace{-12pt}
    \label{fig:Figure12}
\end{figure*}

This section uses ResNet34 as a representative model to analyze the system-level energy consumption breakdown of the VARA architecture, with particular emphasis on evaluating the additional overhead incurred by the IU introduced to support the CAR scheme. Fig.~\ref{fig:Figure12} presents the energy consumption shares of individual modules under VA-Step and VA-ReLU, respectively.

\paragraph{Evaluation Results of VA-Step}
As shown in Fig.~\ref{fig:Figure12}(a), the system energy consumption is primarily contributed by the ADCs and buffers, which account for 61.5\% and 19.0\% of the total, respectively, summing to 80.5\%. The accumulation unit and other peripheral circuits consume 9.2\% and 8.6\%, respectively, while the crossbar contributes only 1.46\%. In contrast, the IU accounts for a mere 0.22\%, substantially lower than all other hardware modules. Therefore, under the VA-Step configuration, the energy consumption of VARA remains dominated by analog-to-digital conversion and data access operations, and the index-access overhead introduced by the IU is virtually negligible within the overall system energy budget.

\paragraph{Evaluation Results of VA-ReLU}
As shown in Fig.~\ref{fig:Figure12}(b), the ADC remains the predominant contributor to system energy consumption, accounting for 51.8\%. The energy shares of the buffers and the accumulation unit rise to 25.2\% and 12.0\%, respectively, representing a notable increase over the VA-Step case. The underlying reason is that VA-ReLU retains multi-bit continuous activation values, which enlarges the bit-width for data storage and transfer and consequently increases the overhead of partial-sum accumulation. The other peripheral circuits and the crossbar account for 9.7\% and 0.99\%, respectively. Notably, the IU consumes a mere 0.26\% of the total, indicating that the index-access overhead introduced by CAR is virtually negligible within the overall system energy budget.

In summary, the energy share of the IU consistently remains below 0.3\%. The underlying reason for this ultralow overhead is that the IU stores not the full activation matrices but rather the permutation indices corresponding to each layer's activation matrix, whose storage capacity is determined solely by the activation row dimension and does not scale linearly with the total number of matrix elements. During inference, the system needs to read the permutation sequence only once prior to the computation of each layer and remap the read addresses of the activation buffer accordingly, without performing explicit data movement on the activation matrices. Therefore, CAR achieves activation matrix reordering with extremely lightweight hardware overhead, and the introduction of the IU does not offset the energy savings gained by CAR through eliminating redundant crossbar computations.

\subsection{Related Work and Comparison}
\label{subsec:related_comparison}
\subsubsection{Related Work}

To more clearly position this work relative to existing studies, Table~\ref{tab:sparse_comparison} summarizes existing ReRAM-based sparsity exploitation schemes. As shown in Table~\ref{tab:sparse_comparison}, prior studies have primarily exploited model sparsity through value sparsity enhancement, bit sparsity enhancement, matrix reordering, and runtime computation skipping. However, these methods differ substantially in the optimization stages they address and the hardware granularities at which sparsity is exploited. Specifically, CAP~\cite{liang2018crossbar}, PIM-Prune~\cite{9218523}, and FORMS~\cite{yuan2021forms} primarily increase the proportion of zero-valued weights through pruning, polarization, or quantization to enhance element-wise sparsity. Nevertheless, these methods focus solely on increasing the number of zero-valued elements and lack a structured reordering mechanism that aggregates scattered zeros into contiguous zero-valued regions. Consequently, their computation-skipping granularity is limited to crossbar rows or columns. In contrast, SME~\cite{9643646}, BitS-Net~\cite{9699384}, and ERA-BS~\cite{10177200} focus on increasing the proportion of zero bits in the binary representations of weights or activations and reduce computational overhead through bit slicing, sparse encoding, or low-order bit skipping. Although these methods effectively reduce bit-level operations, their optimizations remain confined to the binary representation of data. Consequently, they cannot readily create all-zero regions, making it impossible to skip entire ReRAM crossbars.

To improve the efficiency of exploiting unstructured sparsity in hardware, EZC~\cite{11279461}, SME~\cite{9643646}, and SRE~\cite{yang2019sparse} employ row/column compression and local dynamic reorganization to exclude all-zero rows or columns within operation units (OUs) from the computational mapping, thereby reducing redundant operations. However, these methods neither actively increase the proportion of zero-valued elements nor provide structured reordering across the global mapping. Consequently, their computation-skipping granularity remains limited to local rows or columns within a crossbar. CompRRAE~\cite{3287640} employs runtime activation-range estimation to terminate the remaining MAC operations whose contributions to the final output are limited. However, it neither modifies the activation distribution nor reorders the activation matrix. RapPIM~\cite{10323699} identifies and prunes insensitive activation rows or columns based on the number of zeros they contain, while simultaneously skipping partial low-bit computations. This method exploits the existing activation sparsity of the model but neither explicitly increases the ZAR nor reorders the activation matrix. Therefore, its computation-skipping granularity is limited to activation rows or columns. In summary, existing studies typically address only a subset of the stages involved in sparsity enhancement, matrix reordering, and computation skipping. They therefore struggle to simultaneously increase the ZAR and organize scattered zero-valued elements into structured redundancy aligned with the physical structure of ReRAM, thereby limiting the opportunities for crossbar-level computation skipping.

\begin{table*}[t]
  \centering
  \caption{Comparison of VARA with Existing Sparsity Exploitation Schemes for ReRAM-Based Accelerators}
  \vspace{-3pt}
  \label{tab:sparse_comparison}
  \footnotesize
  \renewcommand{\arraystretch}{1.12}
  \setlength{\tabcolsep}{2.5pt}

  \begin{tabularx}{\textwidth}{
    >{\raggedright\arraybackslash}p{1.95cm}
    >{\centering\arraybackslash}p{2.35cm}
    >{\centering\arraybackslash}p{2.05cm}
    >{\centering\arraybackslash}p{0.70cm}
    >{\centering\arraybackslash}p{0.70cm}
    >{\centering\arraybackslash}p{0.70cm}
    >{\centering\arraybackslash}p{0.90cm}
    >{\raggedright\arraybackslash}X}

    \toprule
    \makecell[cl]{\textbf{Related Work}} &
    \makecell[c]{\textbf{Sparsity Target}} &
    \makecell[c]{\textbf{Sparsity}\\\textbf{Granularity}} &
    \makecell[c]{\textbf{BSE}} &
    \makecell[c]{\textbf{ESE}} &
    \makecell[c]{\textbf{MR}} &
    \makecell[c]{\textbf{GCR}} &
    \makecell[tl]{\textbf{Skipping Granularity}} \\
    \midrule

    EZC~\cite{11279461} &
    Weight &
    Value-level &
    \xmark &
    \xmark &
    \cmark &
    \xmark &
    OU-row \\

    CAP~\cite{liang2018crossbar} &
    Weight &
    Value-level &
    \xmark &
    \cmark &
    \xmark &
    \xmark &
    Crossbar/column \\

    PIM-Prune~\cite{9218523} &
    Weight &
    Value-level &
    \xmark &
    \cmark &
    \xmark &
    \xmark &
    Crossbar-row/column \\

    SME~\cite{9643646} &
    Weight &
    Bit-level &
    \cmark &
    \xmark &
    \cmark &
    \xmark &
    Bit-sliced crossbar \\

    SRE~\cite{yang2019sparse} &
    Weight/Activation &
    Value-level &
    \xmark &
    \xmark &
    \cmark &
    \xmark &
    OU-row/column \\

    FORMS~\cite{yuan2021forms} &
    Weight/Activation &
    Value-level &
    \xmark &
    \cmark &
    \xmark &
    \xmark &
    Segment \\

    BitS-Net~\cite{9699384} &
    Weight/Activation &
    Bit-level &
    \cmark &
    \xmark &
    \xmark &
    \xmark &
    Bit \\

    ERA-BS~\cite{10177200} &
    Weight/Activation &
    Bit-level &
    \cmark &
    \xmark &
    \xmark &
    \xmark &
    Crossbar-row/column \\

    CompRRAE~\cite{3287640} &
    Activation &
    Value-level &
    \xmark &
    \xmark &
    \xmark &
    \xmark &
    MAC operation \\

    RapPIM~\cite{10323699} &
    Activation &
    Value-level &
    \xmark &
    \xmark &
    \xmark &
    \xmark &
    Activation-row/column \\

    \textbf{VARA} &
    \textbf{Activation} &
    \textbf{Value-level} &
    \xmark &
    \cmark &
    \cmark &
    \cmark &
    \textbf{Crossbar} \\

    \bottomrule
  \end{tabularx}

  \vspace{3pt}
  \parbox{\textwidth}{\footnotesize
    BSE: bit-sparsity enhancement;
    ESE: element-sparsity enhancement;
    
    MR: matrix reordering;
    GCR: global clustering-based reordering.
  }
  \vspace{-8pt}
\end{table*}

\subsubsection{Comparison}
\paragraph{Comparison of Bitline Current Modulation via Conductance and Voltage}

Prior work SARA~\cite{he2020towards,he2021saving} reduces bitline current through state-aware training (SAT), which maps a larger fraction of weights to the HRS of ReRAM devices. In contrast, the VAT proposed in this work maps a larger fraction of activations to zero input voltage to reduce bitline current. To compare the energy-saving effects of these two approaches, Table~\ref{tab:vara_comparison} summarizes the relevant results. To exclude the additional energy savings contributed by CAR, we report only the hardware energy consumption of VARA with VAT enabled for comparison with SARA. As shown in Table~\ref{tab:vara_comparison}, for BNNs, SARA reduces crossbar and sense amplifier (SA) energy consumption by 40.53\% and 36.78\%, respectively. In comparison, VARA with VA-Step achieves 76.05\% and 66.98\% reductions in crossbar and ADC energy consumption, respectively, on more complex datasets. Under the 8-bit setting, SARA reduces crossbar and SA energy consumption by 11.67\% and 9.17\%, respectively, whereas VARA with VA-ReLU achieves 54.84\% and 39.82\% reductions in crossbar and ADC energy consumption, respectively.
Moreover, since SARA does not incorporate crossbar-level computation skipping, its energy savings are confined to the crossbar and readout circuitry, limiting further reductions in system-level energy consumption.

The above performance differences can be explained by the current characteristics of ReRAM cells. For a ReRAM cell with conductance $G_i$ under an applied input voltage $V_i$, the cell current is given by $I_i=V_iG_i$. SARA reduces $G_i$ by mapping a larger fraction of weights to the HRS. However, as long as the input voltage remains nonzero, an HRS cell still produces a finite read current. Therefore, this approach can reduce only the conduction current of the ReRAM cell and cannot skip the corresponding computation. In contrast, VARA maps more activations to zero, making the corresponding input voltage $V_i$ zero and eliminating the read current of the associated ReRAM cell. More importantly, when an entire input activation vector is zero, VARA can directly skip the corresponding crossbar-level computation, thereby saving substantial energy in the peripheral circuits. Consequently, VARA reduces energy consumption by both lowering the conduction current of ReRAM cells and enabling crossbar computation skipping, resulting in significantly greater energy savings than those achieved by solely modulating device resistance states.

\begin{table*}[t]
  \centering
  \caption{Comparison of VARA with SARA~\cite{he2020towards,he2021saving} and RapPIM~\cite{10323699}}
  \label{tab:vara_comparison}
  \vspace{-6pt}
  \footnotesize
  \renewcommand{\arraystretch}{1.15}
  \setlength{\tabcolsep}{3.5pt}

  \begin{tabularx}{\textwidth}{
    >{\raggedright\arraybackslash}p{2.20cm}
    >{\centering\arraybackslash}X
    >{\centering\arraybackslash}p{1.15cm}
    >{\centering\arraybackslash}p{1.15cm}
    >{\centering\arraybackslash}p{1.15cm}
    >{\centering\arraybackslash}p{1.15cm}
    >{\centering\arraybackslash}p{1.15cm}}

    \toprule
    \multirow{2}{*}{\makecell[c]{\textbf{Related work}}} &
    \multirow{2}{*}{\makecell[c]{\textbf{Evaluation dataset}}} &
    \multirow{2}{*}{\makecell[c]{\textbf{Accuracy}\\\textbf{change}}} &
    \multicolumn{3}{c}{\textbf{Energy reduction}} &
    \multirow{2}{*}{\makecell[c]{\textbf{EE}\\\textbf{gain}}} \\
    \cmidrule(lr){4-6}
    & & &
    \textbf{Array} &
    \textbf{ADC/SA} &
    \textbf{System} &
    \\
    \midrule

    SARA (BNN + SAT) &
    \makecell[c]{MNIST, SVHN, CIFAR10} &
    $\downarrow 0.38$\% &
    40.53\% &
    36.78\% &
    -- &
    -- \\

    \textbf{VARA (VA-Step)} &
    \makecell[c]{CIFAR10, CIFAR100, Tiny ImageNet} &
    \textbf{$\uparrow 0.28$\%} &
    \textbf{76.05\%} &
    \textbf{66.98\%} &
    \textbf{57.37\%} &
    \textbf{$2.35\times$} \\
    
    \midrule

    SARA (8-bit SLC) &
    \makecell[c]{MNIST, SVHN, CIFAR10} &
    $0.00\%$ &
    11.67\% &
    9.17\% &
    -- &
    -- \\

    \textbf{VARA (VA-ReLU)} &
    \makecell[c]{CIFAR10, CIFAR100, Tiny ImageNet} &
    $\downarrow 1.70$\% &
    \textbf{54.84\%} &
    \textbf{39.82\%} &
    \textbf{27.33\%} &
    \textbf{$1.39\times$} \\
    
    \midrule

    RapPIM (Average) &
    CIFAR10 &
    $\downarrow 1.85$\% &
    -- &
    -- &
    44.82\% &
    -- \\

    \textbf{VARA (Average)} &
    \makecell[c]{CIFAR10, CIFAR100, Tiny ImageNet} &
    \textbf{$\downarrow 0.71$\%} &
    \textbf{74.53\%} &
    \textbf{66.40\%} &
    \textbf{60.12\%} &
    \textbf{$2.68\times$} \\

    \bottomrule
  \end{tabularx}

  \vspace{3pt}
  \parbox{\textwidth}{\small
    EE: energy efficiency.
  }
  \vspace{-12pt}
\end{table*}

\paragraph{Comparison of Computation Skipping by Exploiting Activation Sparsity}

Table~\ref{tab:vara_comparison} summarizes the comparison between VARA and RapPIM~\cite{10323699}, a state-of-the-art (SOTA) scheme for sparse-activation optimization. For this comparison, VARA's hardware performance is evaluated with VAT and CAR jointly enabled. As shown in Table~\ref{tab:vara_comparison}, RapPIM reports a 44.82\% reduction in system energy consumption, whereas VARA achieves a reduction of 60.12\%, exceeding RapPIM by 15.30 percentage points. In terms of model accuracy, the five models evaluated by RapPIM on CIFAR10 incur an average accuracy loss of 1.85\%. By contrast, VARA incurs an average accuracy loss of only 0.71\% on more complex datasets (CIFAR10, CIFAR100, and Tiny ImageNet), while achieving a 2.68$\times$ improvement in energy efficiency. These results demonstrate that VARA can effectively translate activation sparsity into system-level energy savings even on more complex datasets.

The above performance differences primarily arise from how the two approaches generate, structure, and exploit sparsity in hardware. As shown in Table~\ref{tab:sparse_comparison}, RapPIM counts the zeros in each row and column of the activation matrix at runtime and prunes activation rows or columns that have little impact on model accuracy. However, RapPIM neither actively increases the ZAR nor structurally reorders the activation matrix. Consequently, the number of rows or columns that can be pruned is constrained by the original activation distribution, while the granularity of computation skipping is largely limited to activation rows or columns. In contrast, VARA jointly optimizes activation sparsity enhancement, structural reorganization, and computation skipping. First, VAT increases the ZAR, creating more zero values for subsequent computation skipping. Building on this enhanced sparsity, CAR clusters scattered, unstructured zeros into contiguous all-zero regions, rendering computations on the corresponding ReRAM crossbars redundant. Therefore, by combining a higher ZAR with coarser-grained computation skipping, VARA achieves significantly greater system-level energy savings than RapPIM.
\section{Conclusion}
\label{sec:section_5}

This paper proposes VARA, a voltage-aware ReRAM accelerator designed to exploit activation sparsity. First, the proposed VAT actively suppresses low-magnitude activations to enhance activation sparsity at the model level. Subsequently, the proposed CAR scheme leverages the co-zero correlation across different activation dimensions to jointly reorder the activation matrix and its corresponding weight mapping, aggregating scattered zeros into contiguous all-zero regions, thereby transforming unstructured element-wise activation sparsity into structured redundancy exploitable by the hardware. Building upon this, VARA reduces system energy consumption by skipping crossbar computations corresponding to all-zero inputs. Experimental results demonstrate that, with only marginal model accuracy degradation, VARA reduces the total system energy consumption by 60.12\% on average and improves the system energy efficiency by 2.68$\times$ on average compared to the baseline, fully validating the effectiveness of the proposed approach.

\bibliographystyle{Utils/ACM-Reference-Format} 
\bibliography{References} 
\end{document}